\documentclass[prd,twocolumn,superscriptaddress,floatfix,nopacs,preprintnumbers,nofootinbib]{revtex4-2}

\usepackage[utf8]{inputenc}
\usepackage{subfigure}
\usepackage[normalem]{ulem}
\usepackage[dvipsnames]{xcolor}
\usepackage{mathrsfs}
\usepackage{amssymb,bm}
\usepackage{amsmath}
\usepackage{mathtools}
\usepackage{slashed}
\usepackage{xfrac}
\usepackage{graphics}
\usepackage{graphicx}
\usepackage{physics}
\usepackage[h]{esvect}
\usepackage{accents}

\renewcommand{\tr}[1]{\mathrm{Tr}\left[#1\right]}

\newcommand{\gev}{\mathrm{GeV}}

\definecolor{lcolor}{rgb}{0.5,0,0}
\definecolor{citcolor}{rgb}{0,0.3,0.0}

\usepackage[breaklinks,colorlinks,citecolor=citcolor,urlcolor=blue,linkcolor=lcolor]{hyperref}
\usepackage[capitalise]{cleveref}

\begin{document}

\author{Dana Avramescu}
\email[Corresponding author: ]{dana.d.avramescu@jyu.fi}
\affiliation{Department of Physics, University of Jyväskylä,  P.O. Box 35, 40014 University of Jyväskylä, Finland}
\affiliation{Helsinki Institute of Physics, P.O. Box 64, 00014 University of Helsinki, Finland}

\author{Vincenzo Greco}
\email{greco@lns.infn.it}
\affiliation{Department of Physics and Astronomy, University of Catania, Via Santa Sofia 64, I-95123 Catania}
\affiliation{INFN-Laboratori Nazionali del Sud, Via Santa Sofia 62, I-95123 Catania, Italy}

\author{Tuomas Lappi}
\email{tuomas.v.v.lappi@jyu.fi}
\affiliation{Department of Physics, University of Jyväskylä,  P.O. Box 35, 40014 University of Jyväskylä, Finland}
\affiliation{Helsinki Institute of Physics, P.O. Box 64, 00014 University of Helsinki, Finland}

\author{Heikki Mäntysaari}
\email{heikki.mantysaari@jyu.fi}
\affiliation{Department of Physics, University of Jyväskylä,  P.O. Box 35, 40014 University of Jyväskylä, Finland}
\affiliation{Helsinki Institute of Physics, P.O. Box 64, 00014 University of Helsinki, Finland}

\author{David M\"{u}ller}
\email{dmueller@hep.itp.tuwien.ac.at}
\affiliation{Institute for Theoretical Physics, TU Wien, Wiedner Hauptstraße 8, A-1040 Vienna, Austria}
\title{The impact of glasma on heavy flavor azimuthal correlations and spectra}

\begin{abstract}
We study the phenomenological impact of the pre-equilibrium glasma initial stage of heavy-ion collisions on heavy quark azimuthal correlations and spectra. Using our numerical solver, we simulate the transport of heavy quark test particles in an SU($3$) glasma background field. The glasma field equations are formulated using classical real-time lattice gauge theory, and the heavy quark dynamics are described by classical transport equations numerically solved using the colored particle-in-cell method. For the first time, the effect of the glasma stage on the azimuthal correlations of $c\overline{c}$ and $b\overline{b}$ pairs is studied. The resulting azimuthal width $\sigma_{\Delta\phi}$ exhibits a large and quick decorrelation due to the strong glasma fields. Further, we evaluate how the $p_T$-broadening in the glasma affects heavy quark $p_T$-spectra, which are initialized according to the Fixed-Order Next-to-Leading Logarithm (FONLL) heavy quark production calculation. The nuclear modification factor $R_{AA}$ is extracted for $c$ and $b$ quarks in the glasma and additional nuclear PDF effects accounting for gluon shadowing are included. 
\end{abstract}

\maketitle 


\section{Introduction}
\label{sec:intro}

Ultra-relativistic heavy-ion collisions (HICs) are experimentally studied with the Large Hadron Collider (LHC) and the Relativistic Heavy Ion Collider (RHIC). They provide the extreme temperature and density conditions necessary to produce the Quark Gluon Plasma (QGP), a medium consisting of deconfined quarks and gluons \cite{Busza:2018rrf,Shuryak:2004cy}. It has been a long-standing difficulty to understand the intricate space-time dynamics of the produced medium. Its properties may not solely be described using directly Quantum Chromodynamics (QCD). The evolution of the bulk medium can be viewed as a stage-by-stage process, where each stage is characterized using an effective model. The standard approach is to use an initial pre-equilibrium stage, followed by a kinetic theory description coupled to relativistic hydrodynamics, which then undergoes particlization and hadronization \cite{Blaizot:1987nc,Heinz:2013th,Heinz:2013wva,Gale:2013da,Kurkela:2015qoa,Gelis:2021zmx}. This description of HICs, together with bayesian analysis \cite{Auvinen:2017fjw,Nijs:2020roc,JETSCAPE:2020mzn}, have been successful at reproducing experimental data and extracting bulk properties of the QGP. An alternative way to infer the features the medium created in HICs is to study hard probes, such as heavy quarks and jets~\cite{vanHees:2005wb,Gossiaux:2008jv,Uphoff:2012gb,Cao:2015hia,Scardina:2017ipo,Prino:2016cni,Dong:2019unq,dEnterria:2009xfs,Apolinario:2022vzg}. Due to their special kinematic regimes, namely large mass for heavy quarks and high energy for jets, they are ``separated'' from the underlying bulk medium and may probe its properties. Additionally, since the hard probes are produced early in the collision, they experience the whole evolution of the QGP and, most intriguingly, can be influenced by the initial stage \cite{Xu:2018gux,Li:2020kax}. The early stage of HIC may be described using the Color Glass Condensate (CGC) \cite{Iancu:2000hn, Iancu:2003xm, Iancu:2005jft, Gelis:2010nm, Gelis:2012ri} effective field theory. Within CGC, high-energy nuclei generate gluon fields which are treated as classical. The collision of CGC nuclei produces the glasma \cite{Kovner:1995ja,Lappi:2006fp,Fukushima:2011nq,Gelis:2012ri}, an over-occupied gluonic medium consisting of strong color fields. The glasma can be expected to have a significant effect on the transport of the hard probes. 

Many recent studies have investigated the impact of the pre-equilibrium stage of HICs on the properties of heavy quarks and jets. The classical transport of heavy quarks in the glasma stage was explored in~\cite{Das:2015aga,Das:2017dsh,Ruggieri:2018rzi,Liu:2019lac,Sun:2019fud,Liu:2020cpj,Khowal:2021zoo,Ruggieri:2022kxv,Pooja:2022ojj,Pooja:2024rnn} and showed to exhibit a different behaviour compared to the diffusion models applicable in the QGP phase \cite{Cao:2013ita,Cao:2015hia,Cao:2018ews,Scardina:2017ipo}. The jet momentum broadening in the glasma was extracted for eikonal jets from glasma lattice field correlators \cite{Ipp:2020mjc,Ipp:2020nfu}, revealing a large jet transport coefficient. Similarly large transport coefficients were obtained using an analytical framework where the glasma fields are treated in the proper time expansion \cite{Carrington:2020sww,Carrington:2021dvw,Carrington:2022bnv,Carrington:2023nty,Carrington:2024hhf}. The heavy quark and jet transport coefficients for hard probes in glasma were extracted from colored particle-in-cell simulations \cite{Avramescu:2023qvv,Avramescu:2023vld} and confirmed the large values previously obtained. 

After the glasma stage, hard probes were studied in pre-equilibrium by using QCD effective kinetic theory (EKT). The values for the heavy quark and jet transport coefficients were shown to be compatible with the large values obtained in the glasma~\cite{Boguslavski:2023alu,Boguslavski:2023fdm,Boguslavski:2023waw}. In~\cite{Du:2023izb}, the heavy quark drag and diffusion coefficients were extracted in the pre-equilibrium stage and in~\cite{Zhou:2024ysb} the thermalization of minijets in an expanding non-equilibrium QGP background was investigated. The heavy quark diffusion coefficient in the infinite mass limit was obtained from lattice field correlators in an over-occupied gluonic plasma resembling the glasma in~\cite{Boguslavski:2018beu,Boguslavski:2020tqz,Boguslavski:2020mzh, Backfried:2024rub}. The heavy quark momentum broadening was extracted for heavy quarks with finite masses from real-time lattice simulations for a highly occupied non-Abelian plasma in~\cite{Pandey:2023dzz}. The impact of anisotropy, a feature inherent to the pre-equilibrium stage, was studied in~\cite{Hauksson:2021okc,Hauksson:2023tze,Kuzmin:2023hko,Barata:2023qds,Barata:2024bqp} for the momentum broadening, quenching and polarization of jets in intermediate stages of HICs.

Energy loss in the glasma was first addressed in~\cite{Aurenche:2012qk} and jet energy loss from synchrotron radiation was evaluated in \cite{Barata:2024xwy} using an analytical description of the glasma electric flux tubes. Beyond the glasma stage, energy loss in the pre-equilibrium phase was investigated using various theoretical descriptions. In \cite{Andres:2019eus}, within the EKRT framework \cite{Eskola:1999fc} which implements a QCD-based initial stage model coupled to hydrodynamics, it was shown that the energy loss in the early stage is suppressed. A variant of the Baier-Dokshitzer-Mueller-Peigné-Schiff and Zakharov (BDMPS-Z) formalism modified to consider emissions in the pre-hydrodynamics stage \cite{Andres:2022bql} was applied to extract the medium-induced radiation and revealed the importance of the early emissions. Recently, an analysis of the in-medium emission rate of soft gluons produced by jets showed the potential sensitivity of jets to the early-stage \cite{Adhya:2024nwx}. Using the Dynamical Radiative and Elastic Energy loss Approach (DREENA) \cite{Zigic:2019sth,Stojku:2020wkh,Ilic:2021ezl}, the effect of the early time dynamics was constrained by $R_{AA}$ and $v_2$ data for high-$p_T$ and was shown to have a moderate contribution. 

The presented literature shows the potential sensitivity of the hard probes on the pre-equilibrium stage. Nevertheless, in the glasma phase, most studies focus on the extraction of either momentum broadening or transport coefficients, which are not experimentally accessible quantities. Thus, the impact of the glasma stage on the dynamics of the hard probes is often implied indirectly. A series of exploratory works investigated the effect of glasma on heavy quark observables. Heavy quark spectra in an SU($2$) glasma were extracted in both AA \cite{Sun:2019fud} and $p$A \cite{Ruggieri:2018rzi,Liu:2019lac} collisions. The resulting nuclear modification factor obtained in $p$A was shown to agree with the experimental $D$ meson $R_{p\mathrm{A}}$ data. Furthermore, in AA it was reported that glasma dynamics generated an initial evolution of $R_{AA}$ with a slope opposite to the standard heavy quark diffusion in the QGP \cite{Das:2017dsh}. Hence, it was suggested that this stage could yet be another mechanism contributing to the $R_{AA}$ and $v_2$ puzzle, which was explored using mostly models applicable in the QGP stage~\cite{Das:2015ana,Xu:2015bbz,Noronha-Hostler:2016eow,Cao:2017umt,Zigic:2018ovr}. However, since the glasma studies were done at a qualitative level, it still remains unclear what is the magnitude of the glasma effect. Hence, it is important to improve the previous models and continue investigating the impact of glasma on heavy quark spectra with additional model improvements. 

The simultaneous description of $R_{AA}$ and $v_2$ poses a challenge to existing theoretical models. It is considered that more differential observables, such as heavy flavor angular correlations \cite{Thomas:2024cso}, might offer better discrimination between various models. Additionally, the angular correlations are considered to be susceptible to the heavy quark production mechanism \cite{Altmann:2024icx}. Inspired by this observation, one may question whether such an observable is potentially sensitive to the initial stage of the collision. Nevertheless, there is no current model calculation of heavy flavor angular correlations in the glasma. A related calculation done in the glasma is the evolution and dissociation of $Q\overline{Q}$ pairs studied in \cite{Pooja:2024rnn}. On the other hand, there are theoretical model estimates for the azimuthal correlations in the QGP. Heavy quark azimuthal angle correlations were extracted in \cite{Nahrgang:2013saa} where the heavy quarks obey the Boltzmann equation with collisional and radiative energy loss effects, and evolve in an ideal hydrodynamic fluid whose initial condition is given by EPOS. Within a similar framework, the azimuthal correlations of heavy quark $Q\overline{Q}$ pairs were studied using EPOS4HQ, across multiple system sizes \cite{Zhao:2024oma}. A simultaneous extraction of $R_{AA}$, $v_2$ and $c\overline{c}$ azimuthal correlations with Langevin dynamics was done in \cite{Cao:2014pka}. Results for the same observables were obtained using the Boltzmann equation with collisions \cite{Scardina:2014lda}. However, these studies lack a treatment of the heavy quark dynamics in the pre-equilibrium stage. The potential sensitivity to the early dynamics motivates the investigation of azimuthal correlations for $Q\overline{Q}$ pairs in the evolving glasma. 

The aim of this study is to investigate the impact of the glasma stage on heavy quark observables, namely heavy flavor two-particle correlations and spectra. Throughout this work, we use our numerical solver for the classical transport of partons in the glasma background fields \cite{Avramescu:2023qvv}. First, we investigate correlations of heavy quarks evolving in the glasma. We start by initializing large ensembles of heavy quark and anti-quark pairs with opposite azimuthal angles and simulating their evolution in the glasma.  Then, we evaluate their two-particle correlations in relative azimuthal angle $\Delta\phi$ and rapidity $\Delta\eta$ and extract the corresponding correlation widths $\sigma_{\Delta\phi}$ and $\sigma_{\Delta\eta}$. We study the time evolution of these correlation widths for charm and beauty quarks, along with the dependence on the glasma saturation momentum $Q_s$ and heavy quark initial $p_T$. Second, we extract the nuclear modification factor $R_{AA}$ in the glasma by studying how $p_T$ momentum broadening in the glasma affects heavy quark $p_T$ spectra. We perform numerical simulations of heavy quarks evolving in an SU($3$) glasma, whose $p_T$ distribution at formation time is given by the Fixed-Order+Next-to-Leading Logarithm (FONLL) heavy quark production calculation in $pp$, and extract $R_{AA}$. We study the proper time evolution, dependence on the glasma saturation momentum $Q_s$ and initial collision energy $\sqrt{s}$, for charm and beauty quarks. Additionally, we improve our setup by initializing heavy quark $p_T$ spectra with a FONLL calculation in AA which incorporates nuclear parton distribution function (nPDF) effects. Lastly, we compare the $R_{AA}$ obtained from numerical glasma simulations with the prediction of an analytical toy model which assumes Gaussian $p_T$ momentum kicks for any initial $p_T$. 

This study begins with Sec.~\ref{sec:method} which contains an overview of the theoretical framework and numerical implementation for the glasma fields and the classical transport of heavy quarks in these fields. In Sec.~\ref{sec:observables}, we outline our procedure to extract the two-particle correlations and the nuclear modification factor. The latter is evaluated from numerical simulations and using an analytical toy model for momentum broadening in glasma. Details about our choice for the numerical parameters and the obtained results are shown in Sec.~\ref{sec:results}, along with a detailed study of parameter dependencies. Lastly, in Sec.~\ref{sec:conclusion} we present concluding remarks and improvements planned for future studies. Additionally, Appendix~\ref{appen:casimirs} includes details about the classical color charge algebra of the SU($3$) group and its effect on the extracted observables.

\section{Framework}
\label{sec:method}

In this section, we present the glasma framework and describe the numerical implementation for the evolution of the glasma fields. Additionally, this section contains an overview of the classical transport of probes in the glasma fields, along with our method to implement numerically the transport equations. More details are discussed in \cite{Avramescu:2023qvv}. 

\subsection{Glasma fields}
\label{subsec:glasma}

The Color Glass Condensate effective theory \cite{Iancu:2000hn, Iancu:2003xm, Iancu:2005jft, Gelis:2010nm, Gelis:2012ri} is based on a kinematic separation between the soft and hard partons within a high-energy nucleus. Since the nuclear wavefunction at high energies is mostly dominated by a large density of gluons \cite{Gyulassy:2004zy}, the soft fields have large occupation numbers and may be treated as classical. In this picture, the hard constituents, mostly valence quarks, are treated as a color current $J^\mu$ that generates the soft gauge fields $A^\mu$ representing the soft gluons.  In this classical approximation, the coupled dynamics between the hard and soft sectors is encompassed in the classical Yang-Mills (CYM) equations 
\begin{equation}
    \label{eq:cym}
    \mathcal{D}_\mu F^{\mu\nu}=J^\nu,
\end{equation}
where $\mathcal{D}_\mu=\partial_\mu-\mathrm{i}g[A_\mu,\,\cdot\,]$ is the covariant derivative, $F^{\mu\nu}=\partial_\mu A_\nu-\partial_\nu A_\mu -\mathrm{i}g[A_\mu,A_\nu]$ the field strength tensor and $g$ the coupling constant. Within the CGC framework, the color current $J^\mu$ generated by the hard partons is approximated using high energy kinematics and a classical probability distribution of color charges. More precisely, one assumes that the charges within nuclei which move along one of the light cone directions\footnote{In our convention, the light cone coordinates are defined as $x^\pm\equiv(x^0\pm x^3)/\sqrt{2}$, while $\vec{x}_\perp\equiv(x^1,x^2)$.} $x^{\pm}$ produce a current
\begin{equation}
    \label{eq:lccurrent}
    J_{A,B}^{\mu,a}(x)=\delta^{\mu\pm}\delta(x^\mp)\rho_{A,B}^a(\vec{x}_\perp),
\end{equation}
where $a$ denotes the color index and $\rho^a$ are  classical color charge densities. One needs additional considerations to describe how classical color charge is distributed within the nucleus. For this purpose, we employ the McLerran-Venugopalan (MV) model \cite{McLerran:1993ni, McLerran:1993ka, McLerran:1994vd}, which is applicable for large nuclei with infinite transverse extent. Inside such nuclei, the color charges are assumed to be Gaussian stochastic variables which obey color neutrality and are uncorrelated in both color and transverse coordinates
\begin{equation}
    \label{eq:mvcharges}
    \begin{aligned}
        &\big\langle \rho^a_{A,B}(\vec{x}_\perp)\big\rangle=0,\\
        &\big\langle \rho^a_{A,B}(\vec{x}_\perp)\rho^b_{A,B}(\vec{y}_\perp)\big\rangle=\delta^{ab}\delta^{(2)}(\vec{x}_\perp-\vec{y}_\perp)g^2\mu^2_{A,B}.
    \end{aligned}
\end{equation}
The MV model parameter $g^2\mu$ is the only physical variable in this ansatz and represents the color charge density inside each nucleus. One can show that  $g^2\mu=\mathcal{O}(Q_s)$, where $Q_s$ is the gluon saturation momentum and is approximately dictated by the collision energy. In numerical glasma simulations which implement the MV model, the proportionality factor between $g^2\mu$ and $Q_s$ depends on the choice of the numerical parameters \cite{Lappi:2007ku, Fukushima:2007ki}. The CYM field equations from Eq.~\eqref{eq:cym} sourced by the light cone color current from Eq.~\eqref{eq:lccurrent} yield an analytic solution for the classical gauge field produced by a single CGC nucleus. In the light cone gauge this solution is given by transverse pure gauge field configurations \cite{Kovner:1995ja,Lappi:2006fp,Lappi:2006hq}
\begin{equation}
    \label{eq:puregauges}
    \alpha^i_{A,B}(\vec{x}_\perp)=\frac{\mathrm{i}}{g}V_{A,B}(\vec{x}_\perp)\partial^i V_{A,B}^\dagger(\vec{x}_\perp),
\end{equation}
with the Wilson lines $V_{A,B}$ computed using the initial MV model color charges.

We proceed to the collision of CGC nuclei whose gauge potentials satisfy Eq.~\eqref{eq:puregauges}. We follow the standard prescription for constructing the resulting glasma fields \cite{Kovner:1995ja,Lappi:2006fp,Fukushima:2011nq,Gelis:2012ri} which relies on the boost-invariant approximation. The glasma fields are parametrized in the Milne coordinates, using the Milne proper time $\tau\equiv\sqrt{2x^+x^-}$ and the space-time rapidity $\eta_s\equiv\ln(x^+/x^-)/2$. In this coordinate system, the boost-invariance is imposed as an $\eta_s$-independence of the gauge fields, namely $A^\mu(x)=A^\mu(\tau, \vec{x}_\perp)$. Additionally, the glasma collision problem is formulated in the technically advantageous temporal gauge $A^\tau=0$. The remaining gauge components are parametrized using the ansatz
\begin{align}
    \label{eq:glansatz}
    \begin{split}
        &A_C^i(\tau,\vec{x}_\perp)=\theta_A \alpha_A^i(\vec{x}_\perp)+\theta_B \alpha_B^i(\vec{x}_\perp)+\theta_C \alpha^i_C(\tau,\vec{x}_\perp),\\
        &A_C^\eta(\tau,\vec{x}_\perp)=\theta_C \alpha^\eta_C(\tau,\vec{x}_\perp),
    \end{split}
\end{align}
with the Heaviside functions $\theta_A\equiv\theta(-x^+)\theta(x^-)$, $\theta_B\equiv\theta(x^+)\theta(-x^-)$ for the light cone domain of the pure gauges generated by nuclei $A,B$, see Eq.~\eqref{eq:puregauges}, and $\theta_C\equiv\theta(x^+)\theta(x^-)$ for the forward light cone where the glasma fields reside. The theta functions assure the correct domains for the gauge fields of each of the colliding nuclei $\alpha_{A,B}$ and the resulting glasma $\alpha^{i,\eta}_C$ field in the forward light cone. Along the light cone, at $\tau\rightarrow 0^+$, the glasma initial condition satisfies the CYM equation of motion
\begin{equation}
    \label{eq:collcym}
    \mathcal{D}_\mu F^{\mu\nu}_C\Big|_{\tau\rightarrow 0^+}=J^\nu_C,\quad\text{with}\quad J^\mu_C=J^\mu_A+J^\mu_B,
\end{equation}
where $J^\mu_{A,B}$ are given by Eq.~\eqref{eq:lccurrent}. The ansatz from Eq.~\eqref{eq:glansatz} together with the CYM equation of motion~\eqref{eq:collcym} evaluated along the boundary of the light cone provide the initial condition
\begin{equation}
    \label{eq:glasmaic}
    \begin{aligned}
        &\alpha^i_C\left(\tau, \vec{x}_{\perp}\right)\Big|_{\tau\rightarrow 0^+}=\alpha_A^{i}\left(\vec{x}_{\perp}\right)+\alpha_B^{i}\left(\vec{x}_{\perp}\right),\\ 
        &\alpha^\eta_C\left(\tau, \vec{x}_{\perp}\right)\Big|_{\tau\rightarrow 0^+}=\frac{\mathrm{i}g}{2}\left[\alpha_A^{i}\left(\vec{x}_{\perp}\right), \alpha_B^{i}\left(\vec{x}_{\perp}\right)\right],
    \end{aligned}
\end{equation}
expressible solely from the pure gauge fields given in Eq.~\eqref{eq:puregauges}. The corresponding initial glasma chromo-electromagnetic fields are purely longitudinal \cite{Lappi:2006fp,Fujii:2008dd,Chen:2013ksa}. 

The expressions from Eq.~\eqref{eq:glasmaic} are used to initialize the glasma fields at $\tau\rightarrow 0^+$. The proper time evolution at $\tau>0$ is then given by the sourceless CYM equations. In Milne coordinates, the field equations read 
\cite{Kovner:1995ja,Fries:2007iy}
\begin{equation}
    \label{eq:cymmilne}
    \begin{aligned}
        \frac{1}{\tau^3} \partial_\tau \tau^3 \partial_\tau \alpha^\eta-\left[\mathcal{D}^i,\left[\mathcal{D}^i, \alpha^\eta\right]\right] & =0, \\ 
        \frac{1}{\tau}\left[\mathcal{D}^i, \partial_\tau \alpha^i\right]-\mathrm{i} g \tau\left[\alpha^\eta, \partial_\tau \alpha^\eta\right] & =0, \\ 
        \frac{1}{\tau} \partial_\tau \tau \partial_\tau \alpha^i-\mathrm{i} g \tau^2\left[\alpha^\eta,\left[\mathcal{D}^i, \alpha^\eta\right]\right]-\left[\mathcal{D}^j, F^{j i}\right] & =0,
    \end{aligned}
\end{equation}
and can be further solved numerically. 

To assure a numerically gauge invariant discretization, we use a real-time lattice gauge theory formulation of these equations \cite{Krasnitz:1998ns,Lappi:2003bi,Lappi:2004mbt}. Due to boost-invariance, $\alpha^\eta$ behaves as a scalar under the $\eta_s$-independent non-Abelian gauge transformation. The transverse gauge fields $\alpha^i$ are replaced by gauge links, which represent the shortest Wilson lines on the lattice\footnote{Here $a=L/N$ is the lattice spacing of a square transverse lattice with length $L$ and $N\times N$ lattice sites, $\boldsymbol{x}\equiv (x_i, y_i)$ is the transverse coordinate of a lattice site $i\in\{1,\dots,N\}$ and $\hat{\imath}$ is the unit vector along $i\in{x,y}$.}
\begin{equation}
    \label{eq:gaugelink}
    \begin{split}                 
        \alpha^i(\tau,\vec{x}_\perp)\xmapsto{\text{lattice}}U_{\boldsymbol{x},\hat{\imath}}(\tau)\approx \exp\left\{\mathrm{i}g a \alpha^i\left(\tau, \vec{x}_\perp+\frac{a}{2}\hat{\imath}\right)\right\}.
    \end{split}
\end{equation}
Similarly, the corresponding field strength tensor $F^{ij}$ is represented by a plaquette variable, which constitutes the simplest Wilson loop constructed on the lattice
\begin{align}
    \label{eq:plaquette}
    \begin{split}
        F^{ij}(\tau,\vec{x}_\perp)&\xmapsto{\text{lattice}}U_{\boldsymbol{x},\hat{\imath}\hat{\jmath}}(\tau)\\
        &\approx\exp\left\{\mathrm{i}g a^2 F^{ij}\left(\tau, \vec{x}_\perp+\frac{a}{2}\hat{\imath}+\frac{a}{2}\hat{\jmath}\right)\right\}.
    \end{split}
\end{align}
The CYM evolution for the glasma fields from Eqs.~\eqref{eq:cymmilne} is recast as a set of partial differential equations for the proper time $\tau$ evolution of $\alpha^\eta_{\boldsymbol{x}}(\tau), U_{\boldsymbol{x},\hat{\imath}}(\tau),$ and $ U_{\boldsymbol{x},\hat{\imath}\hat{\jmath}}(\tau)$ in each lattice site $\boldsymbol{x}$. We solve these numerically using the leapfrog method \cite{Krasnitz:1998ns,Lappi:2003bi,Lappi:2004mbt} implemented in the \texttt{curraun} solver\footnote{Publicly available at \href{https://gitlab.com/openpixi/curraun}{gitlab.com/openpixi/curraun}.}. More technical details on the lattice discretization and numerical implementation of this solver are provided in \cite{Muller:2019bwd, Avramescu:2023qvv}. 

\subsection{Test particles}
\label{subsec:particles}

The classical transport of probes in a Yang-Mills background field is described by Wong's equations \cite{Wong:1970fu,Boozer_2011}. In Milne coordinates, they are expressible as
\begin{subequations}
    \label{eq:wongcurv}
    \begin{align}
        &\frac{\mathrm{d}x^\mu}{\mathrm{d}\tau}=\frac{p^\mu}{p^\tau}, \label{eq:wongcoord}\\
        &\frac{\mathrm{D}p^\mu}{\mathrm{d}\tau}=\frac{g}{T_R}\tr{QF^{\mu\nu}}\frac{p_\nu}{p^\tau}, \label{eq:wongmom}\\ 
        &\frac{\mathrm{d}Q}{\mathrm{d}\tau}=-\mathrm{i}g[A_\mu,Q]\frac{p^\mu}{p^\tau}, \label{eq:wongcharge}
    \end{align}
\end{subequations}
where $(p^\tau)^2\equiv(p^x)^2+(p^y)^2+\tau^2 (p^\eta)^2+m^2$, $m$ is the test particle mass,  $\mathrm{D}/\mathrm{d}\tau$ denotes the covariant derivative in curvilinear coordinates and $T_R$ is the coefficient in $\tr{T^aT^b}=T_R\delta^{ab}$. 
Wong's equations represent the equations of motion for classical test particles probing a Boltzmann-Vlasov distribution function \cite{Heinz:1983nx,Kelly:1994dh,Litim:1999id,Litim:1999ns}. By solving Wong's equations, we effectively sample the distribution function of a collisionless Vlasov kinetic plasma. There is no energy loss mechanism included in this formalism, neither by collisions with other particles nor from classical bremsstrahlung. The momentum evolution from Eq.~\eqref{eq:wongmom} is driven by the color Lorentz force in curvilinear coordinates $\mathcal{F}_\mu\equiv F_{\mu\nu}p^\nu/p^\tau$. Thus, the classical particles experience momentum deflections  due to the background chromo-electromagnetic fields. The classical color charge evolution from Eq.~\eqref{eq:wongcharge} may be recast as a color rotation\footnote{Inspired by the Colored Particle-in-Cell (CPIC) method for classical simulations of particles in non-Abelian fields \cite{Hu:1996sf,Moore:1997sn,Dumitru:2006pz,Dumitru:2005hj,Schenke:2008gg}.}
\begin{equation}
    \label{eq:colorrot}
    Q(\tau)=\mathcal{U}^\dagger(\tau_0,\tau)\,Q(\tau_0)\,\mathcal{U}(\tau_0,\tau),
\end{equation}
where the particle Wilson line accumulates the background gauge field along its trajectory as
\begin{equation}
    \label{eq:partwilsonline}
     \mathcal{U}^\dagger(\tau_0,\tau)=\mathcal{P}\exp\Bigg(-\mathrm{i}g\int\limits_{x^\mu(\tau_0)}^{x^\mu(\tau)}\mathrm{d}x^\mu A_\mu\left(x^\mu(\tau)\right)\Bigg).
\end{equation} 

The color charges are chosen as elements of the classical Lie algebra $\mathfrak{su}(3)$. The equations of motion conserve the values of the classical quadratic $q_2(R)$ and cubic $q_3(R)$  Casimirs\footnote{Defined in analogy with the group theory Casimir invariants \cite{Haber:2019sgz} $T^aT^a\equiv C_2(R)\mathcal{I}_R$ and $d_{abc}T^aT^bT^c\equiv C_3(R)\mathcal{I}_R$, where $\mathcal{I}$ is the identity matrix in the given representation $R$.} 
\begin{equation}
    \label{eq:q23}
    Q^aQ^a\equiv q_2(R),\quad d_{abc}Q^aQ^bQ^c\equiv q_3(R),
\end{equation}
in the representation $R$.
Here, $Q^a$ are the (representation-dependent) color components of the color charge $Q = Q^a T^a$. We are mainly interested in the fundamental $R=F$ representation for quarks. We generate random initial color charges $Q(\tau_0)$ with fixed values for $q_{2,3}$, which then remain conserved under color rotations according to Eq.~\eqref{eq:colorrot}. 

In the classical transport formalism, particles "pick up" contributions from the glasma fields and experience coordinate and momentum kicks, resulting in non-eikonal trajectories. Effectively, the color rotation of the heavy quarks in the glasma fields are performed using finite-$v$ Wilson lines $W_v(x_\perp,\tau)=\mathcal{P}\exp\left[-\mathrm{i}g\int\mathrm{d}\tau^\prime v^\mu A_\mu (x_\perp(\tau^\prime),\tau^\prime)\right]$ numerically evaluated along the particle trajectory. In the glasma, the color rotation from Eq. (12) is done with the Wilson line from Eq. (13), which is a finite-$v$ Wilson line. More details on how this Wilson line is used to numerically apply color rotations along the quark's trajectory are provided in \cite{Avramescu:2023qvv}.

We simulate point particles at continuous spatial coordinates $(\vec{x}_\perp, \eta_s)$ according to the coordinate evolution, Eq.~\eqref{eq:wongcoord}, in a gauge field that is discretized on a lattice in the transverse coordinate.  The equation for the Lorentz force, Eq.~\eqref{eq:wongmom}, thus requires associating a discrete lattice point to a continuous coordinate. For this
we use the nearest grid point (NGP) approximation $\vec{x}_\perp(\tau_n)\approx\mathrm{NGP}(\vec{x}_\perp(\tau_n))=\boldsymbol{x}_n$ which replaces the particle transverse coordinates with the closest lattice site $\boldsymbol{x}_n\equiv\boldsymbol{x}(\tau_n)$. No approximations are required for the $\eta_s(\tau_n)$ rapidity coordinate. The particle Wilson line from Eq.~\eqref{eq:partwilsonline} is numerically approximated as a product of infinitesimal timesteps as $\mathcal{U}(\tau_0, \tau_m)=\prod_n \mathcal{U}(\tau_{n-1}, \tau_n)$, with $n\in\{1, m\}$. Each of the infinitesimal Wilson lines involved in the product simplifies to a product of Wilson lines in the different coordinate directions 
\begin{equation}
    \label{eq:numwilline}
    \begin{aligned}
        &\mathcal{U}(\tau_{n-1},\tau_n) \simeq \exp\Bigg(\mathrm{i}g \int\limits_{\boldsymbol{x}_{n-1}}^{\boldsymbol{x}_n}\mathrm{d}y^{i} A_i\left(\boldsymbol{y}\right)\Bigg)\\
        & \times \exp\left(\mathrm{i}g \delta\eta_n A_\eta(\boldsymbol{x}_n)\right)= U_{\boldsymbol{x}_{n-1},\hat{\imath}}(\tau_n) U_{\boldsymbol{x}_{n},\hat{\eta}}(\tau_n).
    \end{aligned}
\end{equation}
The infinitesimal particle color charge rotation is performed using the transverse gauge link variables $U_{\boldsymbol{x},\hat{\imath}}$ and a rapidity gauge link $U_{\boldsymbol{x},\hat{\eta}}$ constructed from $A_{\boldsymbol{x},\eta}$. This method of discretizing the color rotations has the advantage that, by construction, $Q\in\mathfrak{su}(N_c)$ throughout the evolution and the Casimir invariants from Eq.~\eqref{eq:q23} are numerically conserved for any $Q(\tau)$. We use the following method to randomly generate classical color charges with desired values for the Casimir invariants in Eq.~\eqref{eq:q23}. We start from a handpicked initial $Q_0$ which satisfies the Casimir constraints $q_{2,3}(Q_0)$ and randomize it by performing random $U\in\mathrm{SU}(N_c)$ rotations as $Q(\tau_0)=U Q_0 U^\dagger$. This initial charge is used as input in Eq.~\eqref{eq:colorrot} and is then rotated at each time step with the numerical particle Wilson line from Eq.~\eqref{eq:numwilline}. Using these procedures, we perform numerical color rotations in an SU($3$) gauge invariant manner.

These features are incorporated in our numerical solver \cite{zenodo} for Wong's equations in the glasma background fields\footnote{Available at \href{https://github.com/avramescudana/curraun/tree/wong}{github.com/avramescudana/curraun/tree/wong}.}, which was developed and tested in our previous work \cite{Avramescu:2023qvv}.

The classical transport equations in the kinematic regime $M, p_T \sim Q_s$ neglect effects such as energy loss processes, don't capture the complete treatment of multiple scatterings in a gauge field and are affected by the breakdown of the picture of deterministic trajectories. We do not consider the back-reaction from the particles to the fields, which could be a classical energy loss mechanism. Even with this addition, the classical equations of motion won't account for the Landau–Pomeranchuk–Migdal (LPM) or other color coherence effects present in the medium-induced gluon radiation calculations. Additionally, the classical transport framework oversimplifies the scattering between the quark and the color fields by treating the quark as a point-like particle rather than using its full wave function. Lastly, considering deterministic classical paths neglects other quantum interference and medium coherence effects conceptually present in a full path integral approach for propagating quarks in background CGC fields. Such effects are beyond the classical approximation employed in this work. They may be considered by solving the Dirac equation for quarks in the background non-Abelian fields of the glasma, as done in \cite{Gelis:2004jp,Gelis:2005pb,Gelis:2015eua,Gelis:2019dqb,Tanji:2017xiw}. For example, in \cite{Gelis:2004jp,Gelis:2005pb} the main technical difficulty arises from formulating the Dirac equation in Milne coordinates and that the initial condition for the quark wavefunction in $(\tau,\eta)$ coordinates at $\tau=0$ is non-trivial. Additionally, this approach also neglects the back-reaction from the produced quarks on the gluon color fields. Due to these shortcomings, we focus on the classical transport approach, which considerably simplifies quantifying the effect of the glasma on the quark transport.

Classical radiation for $v\sim 1$ particles may be considered by incorporating the back-reaction from the particles to the fields. Quarks moving with finite $v^\mu$ generate a current $j^\mu$ which couples to the glasma fields in the CYM equation $\mathcal{D}_\mu F^{\mu\nu}=j^\nu$. Nevertheless, including the current for fast charges $v\sim 1$ in non-Abelian CYM numerical simulations gives rise to the Cherenkov instability. This may be fixed in colored particle-in-cell (CPIC) simulations by using implicit numerical schemes for discretizing the field equations of motion \cite{Ipp:2018hai}. Using this method, the Cherenkov instability is cured only for $v^\mu$ along a single spatial direction but in our case, the particles could experience momentum kicks along any direction. Thus, the method may not be used in our calculation. Alternatively, bremsstrahlung effects could perhaps be handled using Soft Collinear Effective Theory (SCET) to evaluate the quark radiation in the presence of a medium, in this case the glasma.

\section{Computing observables}
\label{sec:observables}

This section outlines the techniques used to calculate the two-particle correlations and the nuclear modification factor for heavy quarks as they evolve in the glasma. We explain the procedure to compute the correlation width of $Q\overline{Q}$ pairs from the two-particle correlation and describe how heavy quark $p_T$ spectra are extracted using either our numerical solver or an analytical toy model.

\subsection{Two-particle correlations}
\label{subsec:twopartcorr}
We simulate the transport of heavy quark $Q$ and anti-quark $\overline{Q}$ pairs in the evolving glasma, initially produced back-to-back in azimuthal angle, as suggested by leading order (LO) pQCD heavy quark production \cite{Andronic:2015wma,Dong:2019unq}. At next-to-leading order (NLO), the $Q\overline{Q}$ pairs are no longer formed back-to-back \cite{Vogt:2018oje,Vogt:2019xmm}. 
However, this being the first study of $Q\overline{Q}$ angular correlations, we consider the back-to-back case to have a more direct understanding of the decorrelation induced by the glasma dynamics.
The coordinates, momenta, and color charges of the quarks in the pair are initialized according to simple toy models.  Each heavy quark of mass $m$ has a fixed initial transverse momentum $p_T(\tau_\mathrm{form})$ and is formed at a proper time given by $\tau_\mathrm{form} = 1/(2m_T)$, where the transverse mass is $m_T^2=m^2+p_T^2(\tau_\mathrm{form})$. The quark and anti-quark have opposing momenta with the same magnitude, namely $\vec{p}_T^{\, Q}(\tau_\mathrm{form})=-\vec{p}_T^{\,\overline{Q}}(\tau_\mathrm{form})$, where $\vec{p}_T\equiv (p^x, p^y)$. In all our particle simulations $p^\eta(\tau_\mathrm{form})=0$. The quarks in the pair are produced at the same position. Their coordinates $x_T(\tau_\mathrm{form})$ are randomly distributed in the transverse simulation region, where $x_T\equiv (x,y)$, while $\eta(\tau_\mathrm{form})=0$.  In the following, we denote by $N_\mathrm{tp}$ the number of test particles in the evolution. The $Q\overline{Q}$ pair is not necessarily color neutral since the main mechanism of heavy quark pair production is via gluon fusion $gg\mapsto Q\overline{Q}$. Additionally, the color charges suffer multiple color rotations in the glasma and are very rapidly decorrelated from  their initial charges. Thus we treat the initial classical color charges of the $Q$ and $\overline{Q}$ as being random and independent of each other. However, one should not expect an effect from the specific initial color distribution due to the fast stochastic color charge evolution.

\begin{figure*}[!hbt]
\includegraphics[width=0.95\textwidth]{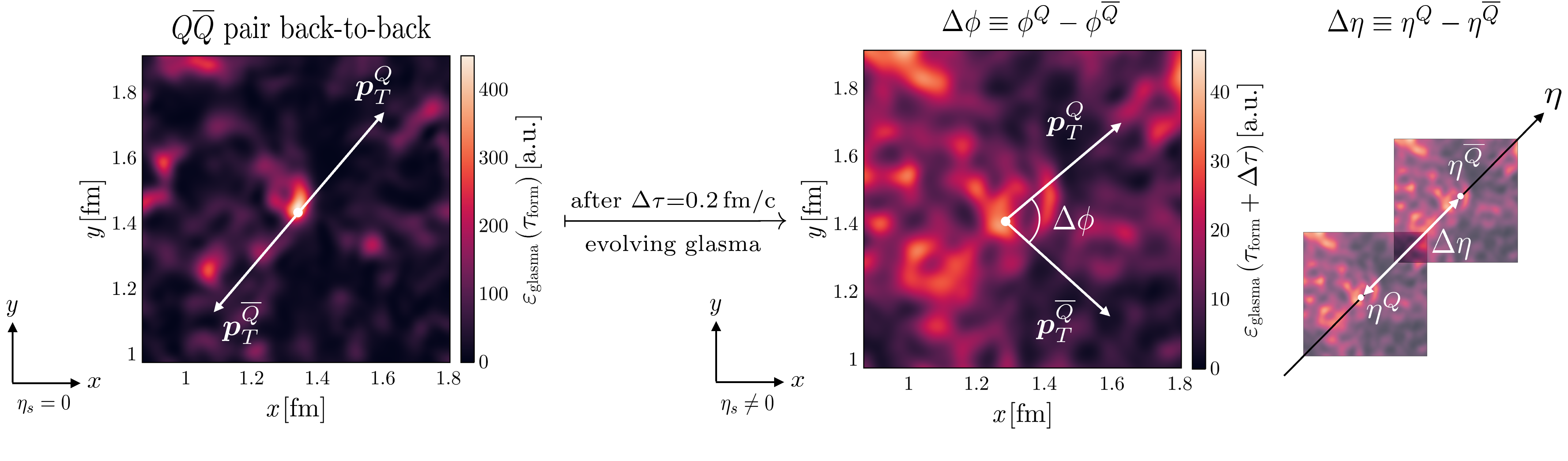}
\caption{\label{fig:sketch_quark_pair} Schematic representation of a single quark ($Q$) anti-quark ($\overline{Q}$) pair evolving in the glasma background field. The quarks in the pair are initialized with opposite $\boldsymbol{p}_T^{Q}=-\boldsymbol{p}_T^{\overline{Q}}$, where $\boldsymbol{p}_T\equiv (p^x, p^y)$. The background depicts a boost invariant slice of the glasma energy density at the formation time of the quarks $\varepsilon(\tau_\mathrm{form})$. After $\Delta\tau$, as the quarks in the pair get deflected by the evolving glasma, their relative azimuthal angle $\Delta\phi\equiv \phi^{Q}-\phi^{\overline{Q}}$ and pseudorapidity $\Delta\eta\equiv\eta^{Q}-\eta^{\overline{Q}}$ shift from the initial peak at $\Delta\phi(\tau_\mathrm{form})=\pi$ and $\Delta\eta(\tau_\mathrm{form})=0$. After $\Delta\tau$, the transverse coordinates of the quarks change are still depicted at the same position even though they numerically change.}
\end{figure*}

We evolve the $Q\overline{Q}$ pair in the glasma background fields using our numerical solver. Wong's equations for coordinates and momenta \eqref{eq:wongmom} in Milne coordinates provide $(x,y,\eta_s)$ and $(p^x, p^y, p^\eta)$ for both the $Q$ and $\overline{Q}$ at each proper time $\tau$. We further extract the laboratory frame longitudinal momentum $p^z=\sinh\eta_s p^\tau+\cosh\eta_s \tau p^\eta$, where $(p^\tau)^2=p_T^2+(\tau p^\eta)^2+m^2$ and $p_T^2=(p^x)^2+(p^y)^2$, along with the pseudorapidity of each quark in the pair as
\begin{equation}
    \label{eq:pseudorapidity}
    \eta^{q}\equiv \frac{1}{2}\ln\frac{p^q+p_z^q}{p^q-p_z^q},\quad\text{where }q\in\{Q,\overline{Q}\},
\end{equation}
with $(p^q)^2=(p_T^q)^2+(p_z^q)^2$ for $q\in\{Q,\overline{Q}\}$. The relative pseudorapidity is given by
\begin{equation}
    \label{eq:deltaeta}
    \Delta\eta\equiv \eta^Q-\eta^{\overline{Q}}.
\end{equation}
We compute the relative azimuthal angle between the quark pair using the scalar and vector product of the transverse momenta of the quarks as
\begin{equation}
    \label{eq:deltaphi}
    \Delta\phi\equiv \arctan\left(\frac{\vec{p}_T^{\,Q}\times \vec{p}_T^{\,\overline{Q}}}{\vec{p}_T^{\,Q}\cdot \vec{p}_T^{\,\overline{Q}}}\cdot \vec{n}\right),
\end{equation}
where $\vec{n}$ is an arbitrary two-dimensional unit vector with $|\vec n|=1$. A sketch showing the evolution and extraction of $\Delta\phi$ and $\Delta\eta$ for a $Q\overline{Q}$ pair overlaid on the glasma energy density profile from a numerical simulation is depicted in Fig.~\ref{fig:sketch_quark_pair}. 

\begin{figure*}[!hbt]
\includegraphics[width=0.9\textwidth]{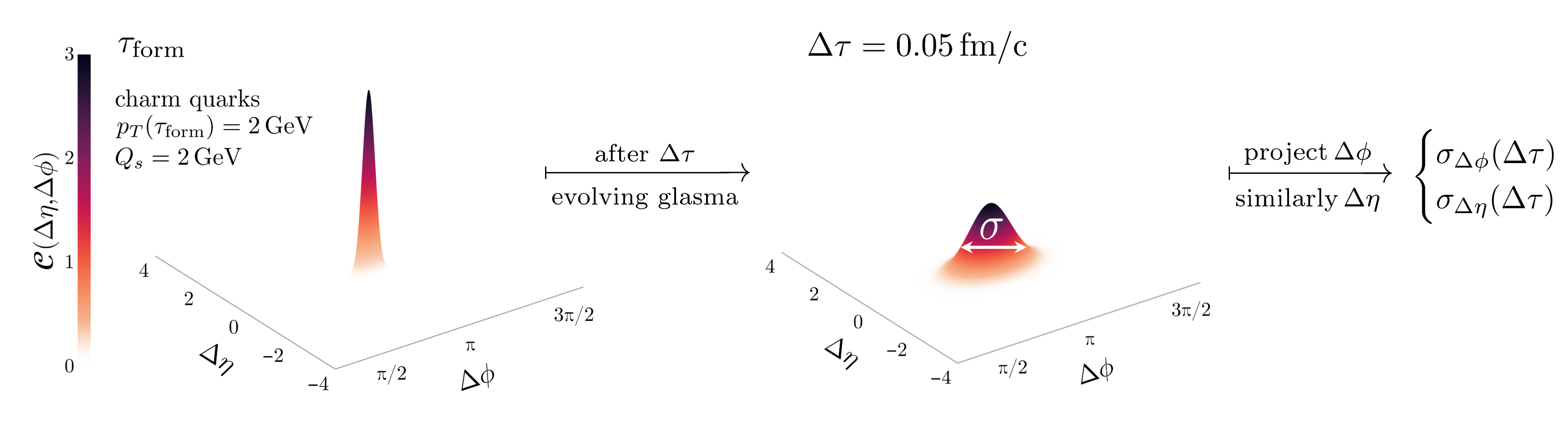}
\caption{\label{fig:sketch_decorrelation} Sketch of the azimuthal and rapidity decorrelation of $Q\overline{Q}$ pairs evolving in the glasma, along with the proper time evolution of the decorrelation widths in $\Delta\phi$ and $\Delta\eta$. The initial two-particle correlation is peaked at $\mathcal{C}(\Delta\phi, \Delta\eta)\propto \delta(\Delta\phi-\pi)\delta(\Delta\eta)$ but after $\Delta\tau$ it gets washed out by the glasma. This is quantified through the decorrelation widths $\sigma_{\Delta\phi}$ and $\sigma_{\Delta\eta}$ of the projected two-particle correlations along $\Delta\phi$ and $\Delta\eta$ respectively, extracted at different proper time values $\Delta\tau$.}
\end{figure*}

After simulating large ensembles of test particles $N_\mathrm{tp}$ in a single glasma event, and a sufficiently large number of glasma events $N_\mathrm{events}$, we collect all the $(\Delta\eta_i,\Delta\phi_i)$ where $i\in\{1,\dots, N_\mathrm{tp}\times N_\mathrm{events}\}$. Then, we extract the two-particle correlations 
\begin{equation}
    \label{eq:twopartcorr}
    \mathcal{C}(\Delta\eta,\Delta\phi)\equiv \frac{1}{N_\mathrm{pairs}}\frac{\mathrm{d}^2N}{\mathrm{d}\Delta\eta\,\mathrm{d}\Delta\phi},
\end{equation} 
and study them as a function of the relative proper time $\Delta\tau\equiv \tau-\tau_\mathrm{form}$, to account for the finite formation time of the quarks in the pair. In practice, we extract these correlations using Kernel Density Estimators (\texttt{KDE}s)\footnote{Package \href{https://github.com/JuliaStats/KernelDensity.jl}{\texttt{KernelDensity.jl}} implemented in \texttt{Julia}.}. By integrating Eq.~(\ref{eq:twopartcorr}) along $\Delta\eta$ or $\Delta\phi$ we also extract the azimuthal and pseudorapidity correlations
\begin{equation}
    \label{eq:cdetacdphi}
    \mathcal{C}(\Delta\phi)\equiv \frac{1}{N_\mathrm{pairs}}\frac{\mathrm{d}N}{\mathrm{d}\Delta\phi},\quad \mathcal{C}(\Delta\eta)\equiv \frac{1}{N_\mathrm{pairs}}\frac{\mathrm{d}N}{\mathrm{d}\Delta\eta}
\end{equation}
as a function of $\Delta\tau$. The momentum kicks in the glasma cause the initial correlation to get smeared out, yielding a symmetric roughly Gaussian distribution in both $\Delta\eta$ and $\Delta\phi$. This is schematically represented in Fig.~\ref{fig:sketch_decorrelation}. To quantify the degree of decorrelation for the initial $Q\overline{Q}$ pair after evolving in the glasma, we further evaluate the $\Delta\tau$ evolution for the widths $\sigma_{\Delta\eta}$ and $\sigma_{\Delta\phi}$ of the correlations $\mathcal{C}(\Delta\eta)$ and $\mathcal{C}(\Delta\phi)$ defined in Eq.~(\ref{eq:cdetacdphi}). 
In practice, we extract the widths $\sigma_{\Delta\eta}$ and $\sigma_{\Delta\phi}$ using the standard deviation of the distributions $\mathcal{C}(\Delta\eta)$ and $\mathcal{C}(\Delta\phi)$.

\subsection{Nuclear modification factor}
\label{subsec:raa}

Classical particles evolving in the glasma experience momentum deflections due to the background electromagnetic fields. The convolution of the probability distribution of deflections with the initial spectrum of quarks leads to a modification of the final $p_T$-spectrum from the initial one. This modification is often quantified by a ratio of the final and initial  $p_T$ spectra. We rely on heavy quark production pQCD calculations \cite{Nason:1989zy,Cacciari:1998it,Beenakker:1990maa,Andronic:2015wma,Helenius:2018uul} to initialize a realistic initial $p_T$-spectrum. We use the FONLL (Fixed Order + Next-to-Leading Logarithm) \cite{Cacciari:1998it,Cacciari:2001td} heavy quark differential cross section $\mathrm{d}\sigma/\mathrm{d}p_T$ to distribute the initial probability $\mathrm{d}N/\mathrm{d}p_T(\tau_\mathrm{form})$ at the formation time of the quark. We study only bare quarks and no fragmentation functions to the subsequent mesons are considered. The FONLL calculation, publicly available \cite{fonll}, is used to initialize heavy quark spectra with both proton and nuclear parton distribution functions (PDFs and nPDFs), at different center-of-mass collision energies $\sqrt{s}$. We denote the choice of the PDF by a superscript ``pp'' or ``AA'' in the following. Moreover, each FONLL heavy quark production cross section comes with factorization and renormalization scales, heavy quark mass and PDF uncertainties. In practice, we approximate the FONLL distribution with a fit performed on a logarithmic scale, which lies within the aforementioned uncertainties 
\begin{equation}
    \label{eq:fonllsigma}
    \frac{\mathrm{d} \sigma^{pp/AA}_{Q\overline{Q}}}{\mathrm{d} p_T}(\sqrt{s}, \mathrm{PDF}/\mathrm{nPDF}) = \dfrac{x_0 p_T}{(1+x_3 p_T^{x_1})^{x_2}}.
\end{equation}
Here the fit parameters $(x_0,x_1,x_2,x_3)$ depend on the initial $\sqrt{s}$, PDF choice for $pp$ or nPDF choice for $AA$ collisions, and quark flavor. Our choices for PDFs are CTEQ6.6~\cite{Nadolsky:2008zw} and the combination CT14NLO~\cite{Dulat:2015mca} for PDF and EPPS16~\cite{Eskola:2016oht} for nPDF. We use the FONLL $p_T$ differential cross section, normalized by the total $p_T$ integrated  cross section, to construct the probability distribution of heavy quarks produced at $\tau_\mathrm{form}$ in the glasma
\begin{equation}
    \label{eq:fonlldndpt}
    \frac{\mathrm{d} N^{pp/AA}}{\mathrm{d} p_T}(\tau_\mathrm{form})= \dfrac{1}{\sigma_{\mathrm{tot},Q\overline{Q}}^{pp/AA}}\frac{\mathrm{d} \sigma^{pp/AA}_{Q\overline{Q}}}{\mathrm{d} p_T},
\end{equation}
where $A$ is the nuclear mass number and $\sigma_{\mathrm{tot},Q\overline{Q}}=\int\mathrm{d}p_T\, \mathrm{d}\sigma_{Q\overline{Q}}/\mathrm{d}p_T$ represents the total $Q\overline{Q}$ cross section but for simplicity we further denote it as $\sigma_{\mathrm{tot}}$. The dependence of the quark probability density on $\sqrt{s}$ and PDF in $pp$ or nPDF in $AA$ is implicit. Note that throughout this work $\mathrm{d}N/\mathrm{d}p_T$ refers to  the probability distribution and not multiplicity, i.e.~$\int \mathrm{d}p_T \, \mathrm{d}N/\mathrm{d}p_T = 1$.

The heavy quark test particles, distributed according to the initial $p_T$ spectrum at formation time $\mathrm{d} N/\mathrm{d} p_T (\tau_\mathrm{form})$ in Eq.~\eqref{eq:fonlldndpt} from the FONLL fit in Eq.~\eqref{eq:fonllsigma}, evolve in the glasma fields and experience $p_T$ momentum broadening. The resulting probability distribution $\mathrm{d} N/\mathrm{d} p_T (\tau)$ in the glasma depends on the evolution time $\tau$. The change in the $p_T$ distribution due to the glasma is encoded in the nuclear modification factor defined as 
\begin{equation}
    \label{eq:raa}
    R_{AA}(\tau)=\dfrac{1}{A^2}\dfrac{\sigma_{\mathrm{tot}}^{AA}}{\sigma_{\mathrm{tot}}^{pp}}\dfrac{\dfrac{\mathrm{d} N}{\mathrm{d} p_T}(\tau;pp/AA)}{\dfrac{\mathrm{d} N^{pp}}{\mathrm{d} p_T}(\tau_\mathrm{form})}.
\end{equation}
Recall that the notation $pp/AA$ used in the glasma spectrum refers to the input FONLL calculation used to initialize the heavy quark distribution from Eq.~\eqref{eq:fonlldndpt}, namely $pp$ with a specified PDF or $AA$ with an nPDF. When a $pp$ FONLL initialization is used, the total cross section is simply $\sigma_{\mathrm{tot}}^{AA}=A^2 \sigma_{\mathrm{tot}}^{pp}$ and the nuclear modification factor $R_{AA}$ from Eq.~\eqref{eq:raa} becomes the ratio of the (normalized) probability distributions for heavy quarks before and after evolving in the glasma. Thus, in this case, $R_{AA}$ must obey a ``sum rule'' where values $R_{AA}>1$ must be compensated by $R_{AA}<1$ at another $p_T$. For the $AA$ FONLL input calculation, due to nuclear PDF effects, $\sigma_{\mathrm{tot}}^{AA}<A^2 \sigma_{\mathrm{tot}}^{pp}$ and consequently $R_{AA}<1$ even without any effects from the glasma. 

In practice, to evaluate $R_{AA}$ from Eq.~\eqref{eq:raa}, we extract heavy quark spectra after their evolution in the glasma phase directly from numerical simulations. We will compare these to the toy model discussed below which assumes Gaussian broadenings at fixed $p_T$. 

\begin{figure*}[!hbt]
\includegraphics[width=0.95\textwidth]{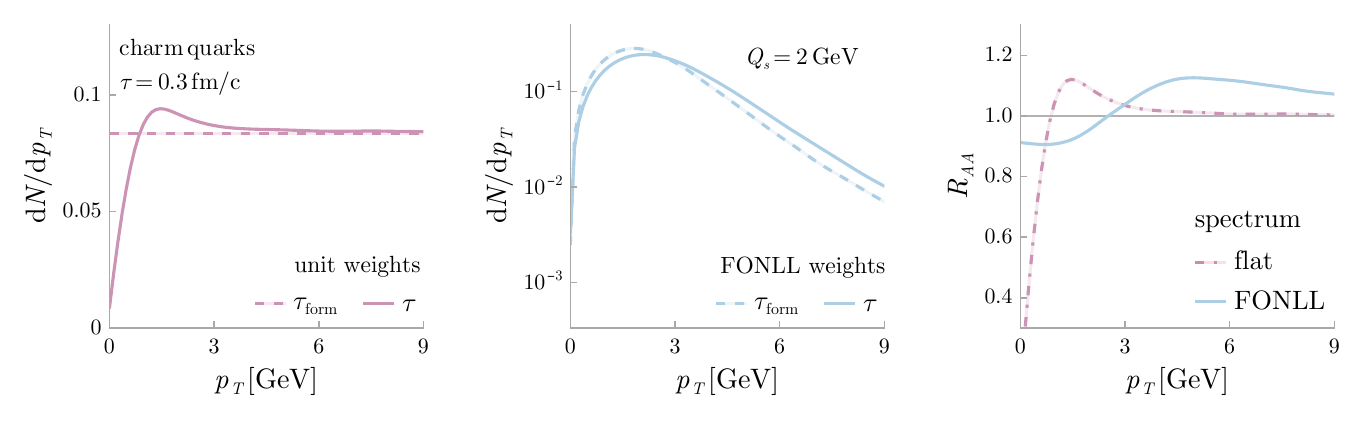}
\caption{\label{fig:sketch_raa_gl_fonll} Heavy quark $p_T$ spectra (\textit{left and middle}) and their corresponding nuclear modification factor $R_{AA}$ (\textit{right}). The \textit{left subfigure} depicts the effect of the glasma at $\tau$ on an initially flat-$p_T$ distribution at $\tau_\mathrm{form}$, while the \textit{middle subfigure} contains the glasma spectra weighted by the initial FONLL distribution.}
\end{figure*}

Even though Wong's equations are applicable for any order in $Q_s/M$, the factorization between the production and propagation in glasma becomes questionable at $Q_s/M\sim 1$. First, the classical quark equations of motion are obtained by taking the classical limit in the colored Dirac equation, which makes no additional assumption about the quarks' $M$ or $p_T$ \cite{Wong:1970fu}. Nevertheless, since we neglect the back-reaction from the quark to the fields by not including the quark current $j^\mu$ in the CYM equations, we are effectively working in the large $M$ limit, appropriate for heavy quarks. Second, the validity of factorizing the production and propagation becomes questionable when $Q_s/M\sim 1$ since the neglected quantum corrections from the CGC field might become relevant during production. In principle, a better approach would be to couple the production and propagation of the quarks in the glasma fields dynamically, e.g. by both considering quark production in strong non-Abelian background fields and solving the Dirac equation in these fields, as done in \cite{Gelis:2004jp,Gelis:2005pb,Gelis:2015eua,Gelis:2019dqb,Tanji:2017xiw}. In practice, overcoming these limitations is cumbersome. In our work, we employ Wong's equations within their appropriate range of applicability and study the leading effect, coming from the classical transport framework, on heavy quark observables.

\subsubsection{Numerical $R_{AA}$ (glasma simulations)}
Similarly to the initialization  for the $Q\overline{Q}$ pairs in Sec.~\ref{subsec:twopartcorr}, we simulate heavy quark test particles with given mass $m$, initial $p_T(\tau_\mathrm{form})$ whose evolution in the glasma starts at $\tau_\mathrm{form}= 1/(2m)$. The quarks are randomly distributed uniformly in transverse coordinate $x_T(\tau_\mathrm{form})$. We initialize the transverse momenta $p_T$ for ensembles of $N_\mathrm{tp}$ test particles according to the FONLL spectra for either $pp$ or $AA$ collisions from Eq.~\eqref{eq:fonllsigma}. At fixed $\tau$ values, the distribution of the quarks is numerically reconstructed, $\mathrm{d} N/\mathrm{d} p_T(\tau)$ is extracted as a \texttt{KDE} and $R_{AA}$ is computed using Eq.~\eqref{eq:raa}. This procedure is repeated for many glasma events $N_\mathrm{events}$. In practice, the sampling from the FONLL spectra is implemented using a weighting procedure. More precisely, we perform the evolution in the glasma of a large ensemble of test particles initialized according to a flat $p_T$-distribution and construct $\mathrm{d} N/\mathrm{d} p_T(\tau)$ as a weighted \texttt{KDE}, where the weights are given by the initial $\mathrm{d} N/\mathrm{d} p_T(\tau_\mathrm{form})$ for each $p_T$ value.  Compared to a straightforward sampling from the steeply falling FONLL spectrum this reweighting procedure reduces the number of glasma events required to obtain enough statistics at high $p_T$. We have also checked  this reweighting method with a direct sampling from the FONLL distribution. This weighting procedure is schematically depicted in Fig.~\ref{fig:sketch_raa_gl_fonll}.

\subsubsection{Analytical $R_{AA}$ (toy model)}
\label{subsec:analtoyraa}

\begin{figure*}[!hbt]
\includegraphics[width=0.9\textwidth]{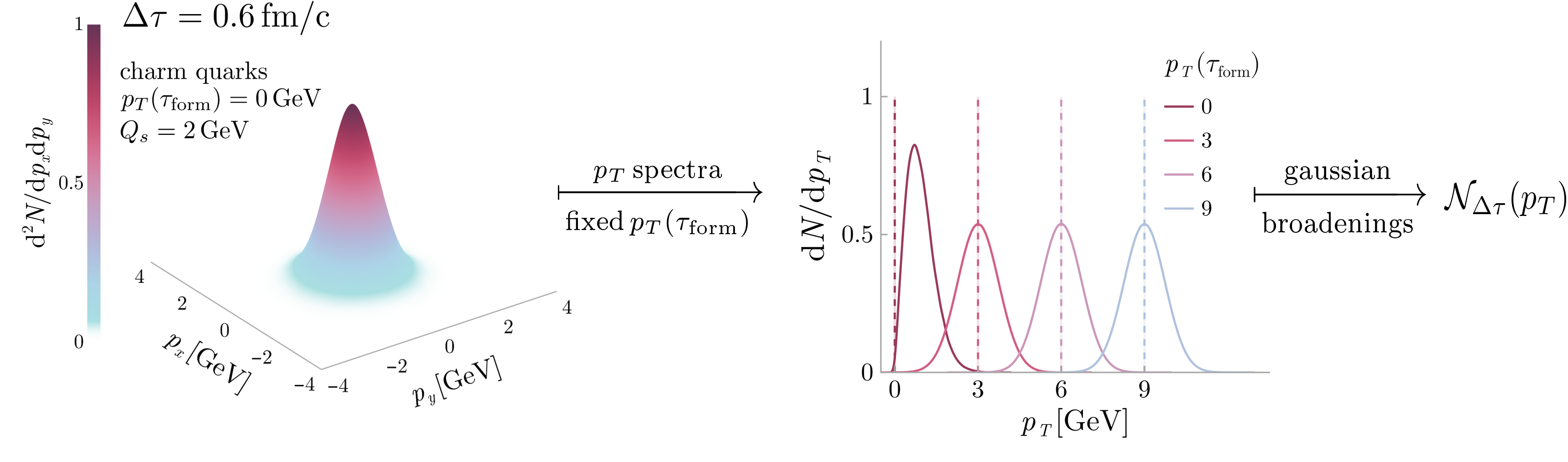}
\caption{\label{fig:sketch_dndpt} 
Broadening in $(p_x, p_y)$ at a fixed $\Delta\tau$ for charm quarks initialized with a given $p_T(\tau_\mathrm{form})$, with various values of initial $p_T(\tau_\mathrm{form})\in\{0,3,6,9\}\,\mathrm{GeV}$. 
In our Gaussian toy model introduced in Sec.~\ref{subsec:analtoyraa}, we approximate this $p_T$-broadening by a Gaussian, which leads to the expression in Eq.~\eqref{eq:raatoyanaly} for the modified momentum distribution.}
\end{figure*}

In addition to the full glasma simulation, we also study a simple toy model for the glasma modification of the single quark spectrum. For this purpose, we assume that the momentum accumulation obeys a two-dimensional Gaussian in $(p_x, p_y)$ as represented in Fig.~\ref{fig:sketch_dndpt}. The width of the Gaussian can then be taken to depend on the proper time $\Delta\tau$.  Thus, the heavy quarks acquire transverse $p_T$ momentum kicks, causing their $p_T$ distribution to change. The effect of such Gaussian momentum kicks on the probability distribution $\mathrm{d}N/\mathrm{d}p_T$ is depicted in Fig.~\ref{fig:sketch_dndpt} for charm quarks initialized with a selection of values for $p_T(\tau_\mathrm{form})$. 

Let us denote the $p_T$ distribution of quarks in the glasma at a certain proper time $\tau$ value as
\begin{equation}
    \label{eq:dndptnot}
    \mathcal{N}_\tau(p_T)\equiv \dfrac{\mathrm{d}N}{\mathrm{d}p_T}(p_T,\tau),
\end{equation}
with $\mathcal{N}_0\equiv \mathcal{N}_{\tau_\mathrm{form}}$ the initial distribution, chosen according to Eq.~\eqref{eq:fonllsigma} for the $pp$ case. Using this notation, the nuclear modification factor defined in Eq.~\eqref{eq:raa} is, in the absence of nPDF effects, expressed as
\begin{equation}
    \label{eq:raatoy}
    R_{AA}(p_T,\tau;\mathcal{N}_0)=\dfrac{\mathcal{N}_\tau(p_T)}{\mathcal{N}_0(p_T)}
\end{equation}
and  $\mathcal{N}_\tau$ implicitly depends on the initial distribution $\mathcal{N}_0$.

In the Gaussian toy model, we assume that the broadening of particle momentum, initialized as $p_T(\tau_0)$, is described by a Gaussian. Moreover, we assume that each $p_T$-bin is affected independently, which corresponds to the fact that there are no interactions and thus no momentum exchange between particles. For a given initial $\vec{q}\equiv (q_x,q_y)$, the momentum exchange is symmetrically centered around the initial momentum $\vec{q}$ and has the same widths $\sigma_x=\sigma_{y}\equiv\sigma$ along the transverse directions
\begin{equation}
    \label{eq:gauss2dxy}
   \mathcal{K}_\tau(\vec{q}, \vec{p})=\dfrac{1}{{2\pi \sigma_{\tau}^2(q_T)}}\exp\left\{-\dfrac{(\vec{p}-\vec{q})^2}{2\sigma_\tau^2(q_T)}\right\}.
\end{equation}
The width of the Gaussian could be taken to be a function of the initial transverse momentum $q_T\equiv \sqrt{q_x^2+q_y^2}$ and the proper time $\tau$, although for simplicity, we assume that it is independent of the initial $p_T$. Different values of the glasma saturation momentum $Q_s$ and the particle quadratic Casimir $q_2$ can be modeled by different values for the width.

The spectrum at $\tau$ is a convolution of the initial spectrum $\mathcal{N}_0$ with all possible Gaussian $p_T$-migrations from Eq.~\eqref{eq:gauss2dxy}, namely
\begin{equation}
    \label{eq:dndptconv}
    \mathcal{N}_\tau(p_T)=\int\mathrm{d}\theta_p \iint\mathrm{d}^2\vec{q}\,\,\mathcal{K}_\tau(\vec{q}, \vec{p})\,\,\mathcal{N}_0(q_T)
\end{equation}
where $\theta_p\equiv \arctan(p_x/p_y)$. After expressing Eq.~\eqref{eq:dndptconv} in polar coordinates and using the integral representation of the modified Bessel function of the first kind $I_0$ \cite{AbramowitzStegun}, the evolved spectrum $\mathcal{N}_\tau$ used to extract the nuclear modification factor from Eq.~\eqref{eq:raatoy} becomes
\begin{align}
    \label{eq:raatoyanaly}
    \begin{split}
            \mathcal{N}_\tau(p_T;\mathcal{N}_0)&=\int\mathrm{d}q_T\,\dfrac{\pi q_T}{{2 \sigma_{\tau}^2(q_T)}}\times\\
            &\phantom{=\int}\exp\left\{-\dfrac{p_T^2+q_T^2}{2\sigma_\tau^2(q_T)}\right\}\mathcal{N}_0(q_T)\,I_0\left(\dfrac{p_T q_T}{\sigma_\tau^2(q_T)}\right).
    \end{split}
\end{align}
Once the width $\sigma_\tau(q_T)$ is fixed and the initial distribution $\mathcal{N}_0$ is known, in our case from the FONLL calculation in Eq.~\eqref{eq:fonllsigma}, the integral from Eq.~\eqref{eq:raatoyanaly} may be performed numerically. The resulting toy model $R_{AA}$ encodes the approximation of perfectly Gaussian momentum broadening in the glasma.

\section{Results}
\label{sec:results}

This section contains an overview of the numerical parameters used as input in our numerical solver, to initialize the glasma fields and the heavy quark transport in these fields. Further, we provide more numerical details about our procedure to extract the two-particle correlations of $Q\overline{Q}$ pairs and the heavy quark nuclear modification factor $R_{AA}$, following their evolution in the glasma, and showcase the key findings, see the companion Letter \cite{Avramescu:2024xts}.

\subsection{Numerical parameters}
\label{subsec:numparams}

We simulate glasma fields obtained in collisions at LHC between lead nuclei, where the value of the saturation momentum in central events roughly corresponds to $Q_s=2\,\mathrm{GeV}$. Since the choice of $Q_s$ is slightly ambiguous, we study the dependence on the initial $Q_s$ of our observables of interest. The relation with the MV model parameter from Eq.~\eqref{eq:mvcharges} is given by $g^2\mu\approx 0.8 Q_s$. In classical Yang-Mills theory, the coupling constant $g^2=4\pi \alpha_s(Q_s)$ can be scaled out from the equations of motion, but the explicit value is needed in order to express quantities such as the energy density into physical units. Its value is taken to run in accordance with the saturation momentum as
\begin{align}
    \alpha_s(Q_s)=\dfrac{1}{\dfrac{33-3 N_{f}}{12 \pi} \ln \dfrac{Q_s^2}{\Lambda_{\mathrm{QCD}}^2}},
\end{align}
where $N_f=3$ and $\Lambda_\mathrm{QCD}=200\,\mathrm{MeV}$. The most important numerical parameters of the lattice are the transverse simulation length which we take to be $L=10\,\mathrm{fm}$ and the number of lattice points $N=512$. In the glasma numerical implementation, the UV cutoffs come from both the lattice spacing $a=L/N$ and the momentum cutoff $\lambda$ used in solving the Poisson equation for the color charges. More details are provided in our earlier study, see \cite{Avramescu:2023qvv}. We test the effect of these UV cutoffs, present in the CYM simulations and inherited by Wong's equations. We validate our results by varying the UV cutoffs and checking that the resulting observables, namely $p_T$-spectra and two-particle correlations, are insensitive to them.

The heavy quark test particles are distributed uniformly in the transverse simulation region $\vec{x}_\perp(\tau_\mathrm{form})\in[0,L]^2$, while their initial space-time rapidity is null $\eta_s(\tau_\mathrm{form})=0$. Each heavy quark is given an initial transverse momentum $p_T(\tau_\mathrm{form})$ and is formed at a fixed initial $\tau_\mathrm{form}$ equal to the inverse of its mass. We study charm and beauty quarks with given masses $m_\mathrm{c}=1.27\,\mathrm{GeV}$ and $m_\mathrm{b}=4.18\,\mathrm{GeV}$ \cite{ParticleDataGroup:2022pth}. Typically, we use a single glasma event for $N_\mathrm{tp}=10^5$ such heavy quark test particles. We simulate multiple glasma events with variable $N_\mathrm{events}$, until convergence for our chosen quantity is numerically achieved. 

Numerically, we sample the color charge components of the heavy quarks such that they satisfy the Casimir constraints from Eq.~\eqref{eq:q23} with $q_2=C_2(F)$ and $q_3=0$, where $C_2(F)=4/3$ is the $\mathrm{SU(3)}$ group theory Casimir in the fundamental representation. More details are provided in Appendix~\ref{subappen:casscal}. This choice is based on our previous observation \cite{Avramescu:2023qvv} that the momentum broadening for partons in certain kinematic scenarios, such as eikonal jets and static quarks, scales with the quadratic Casimir $\langle p^2\rangle_R\propto q_2(R)$ and is independent of the cubic one $q_3(R)$. Inspired by this observation, we numerically check that this scaling holds for $\langle p^2\rangle$ of heavy quarks beyond the infinite mass approximation, with the corresponding change in the width of the azimuthal $\sigma_{\Delta\phi}$ and rapidity correlations $\sigma_{\Delta\eta}$. The results are shown in Appendix~\ref{subappen:cubiccasimirs}. In a similar way, we checked that the $p_T$-spectra of heavy quarks in the glasma are independent of $q_3$.

\subsection{Two-particle correlations}

\begin{figure*}[t]
    \centering
    \includegraphics[width=0.65\columnwidth]
    {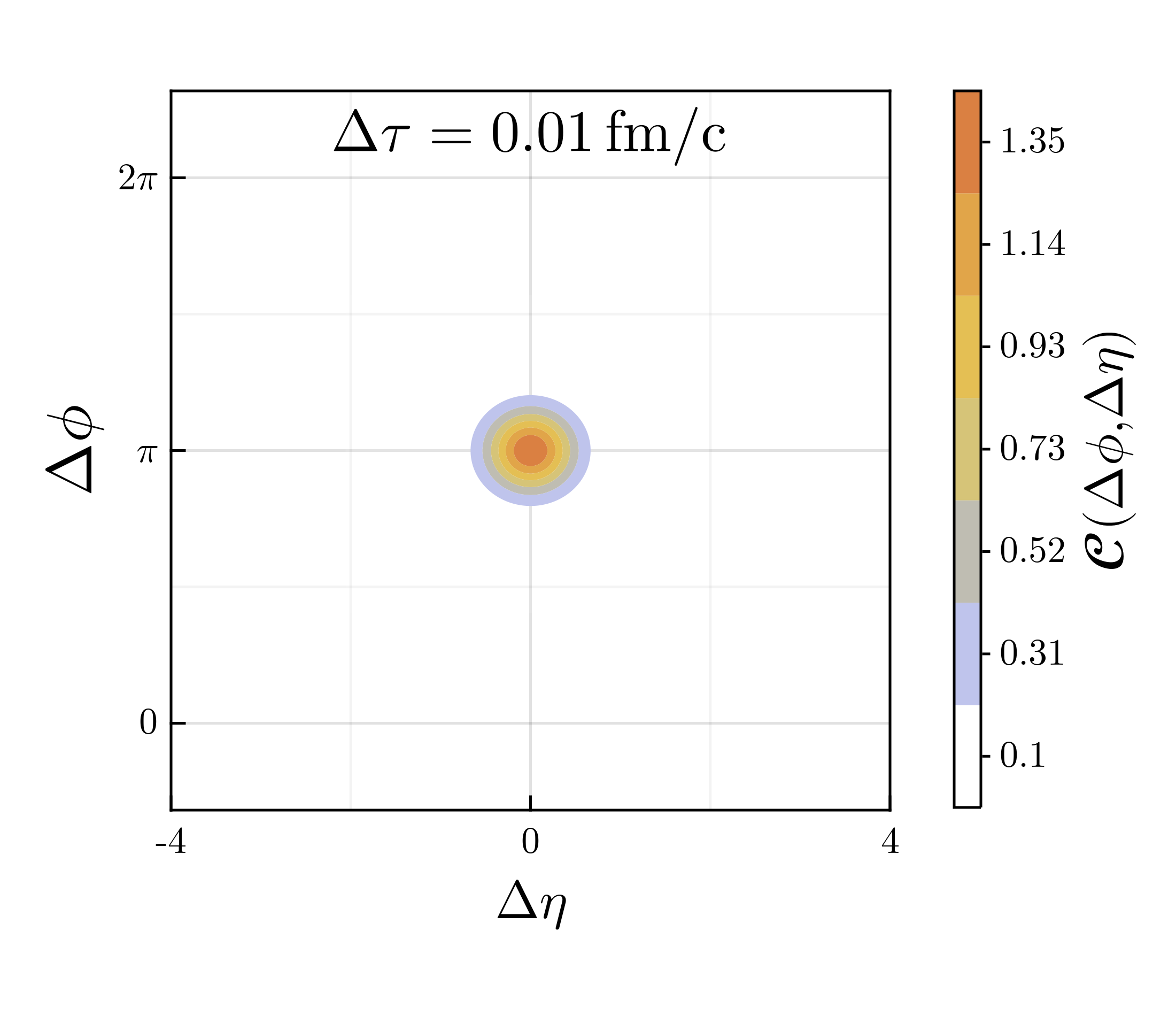}\quad\includegraphics[width=0.65\columnwidth]
    {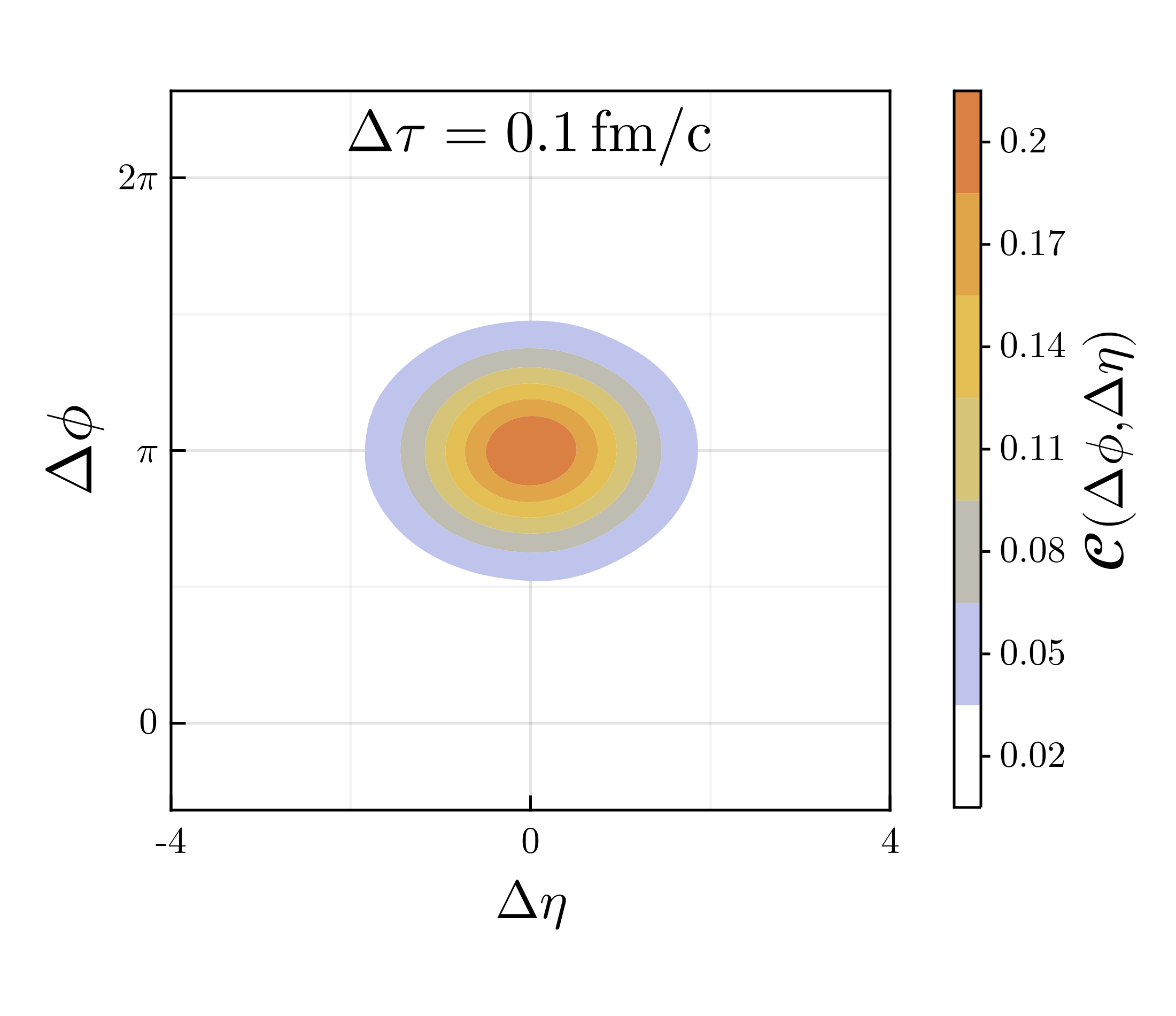}\quad\includegraphics[width=0.65\columnwidth]
    {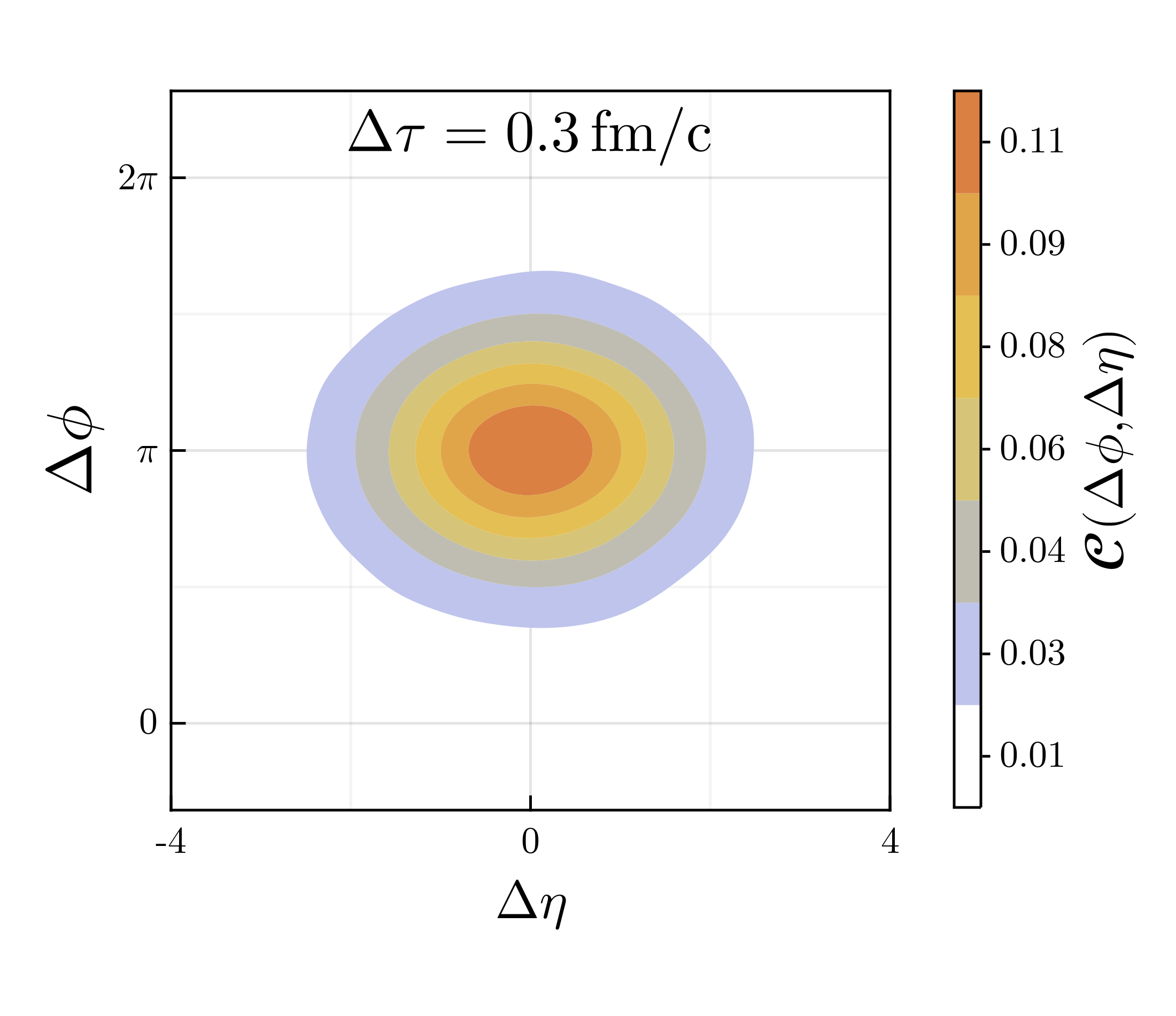}
    \caption{The two-particle correlation $\mathcal{C}(\Delta\eta, \Delta\phi)$, as defined in Eq.(20), at different values of $\Delta\tau$ namely $\Delta\tau\in\{0.01, 0.1, 0.3\}\,\mathrm{fm/c}$, for a single glasma event of $Q_s=2\,\mathrm{GeV}$ with $N_\mathrm{tp}=10^5$ charm test particles initialized with fixed $p_T(\tau_\mathrm{form})=1\,\mathrm{GeV}$. The \textit{left panel} contains the almost initial highly peaked correlation, while the \textit{middle panel} shows decorrelations in both $\Delta\eta$ and $\Delta\phi$. The \textit{right panel} contains the correlation at $\Delta\tau=0.3\,\mathrm{fm/c}$, the typical time at which the glasma stage at LHC ends. The magnitude of the correlation (\textit{color gradient}) is extracted as a \texttt{KDE} and is not normalized.}
    \label{fig:3dtwopartcorr}
\end{figure*}

\begin{figure*}[!hbt]
\includegraphics[width=0.6\textwidth]{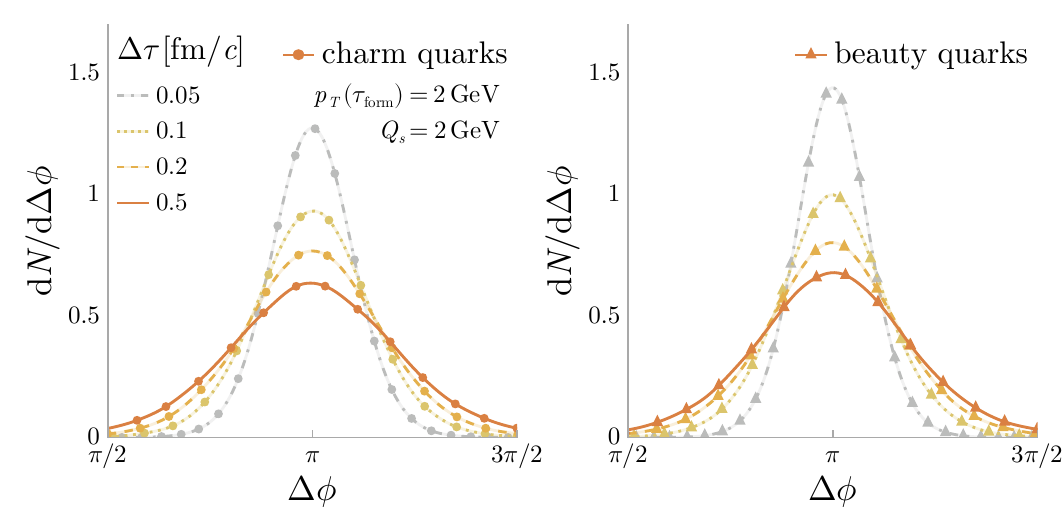}
\caption{\label{fig:dndphi_tau}Azimuthal angle decorrelation $\mathcal{C}(\Delta\phi)$, as expressed in Eq.~\eqref{eq:cdetacdphi}, for charm (\textit{left panel}) and beauty (\textit{right panel}) quarks initialized with $p_T(\tau_\mathrm{form})=2\,\mathrm{GeV}$. The decorrelation is extracted at various values of proper time $\Delta\tau\in\{0.05, 0.1, 0.2, 0.5\}\,\mathrm{fm/c}$ (\textit{different colors and line styles}).}
\end{figure*}

Based on the setup described in Sec.~\ref{subsec:twopartcorr}, we numerically extract the azimuthal and rapidity two-particle correlation $\mathcal{C}(\Delta\eta,\Delta\phi)$ using Eq.~\eqref{eq:twopartcorr} for pairs of charm or beauty quarks and their corresponding antiquarks, throughout their evolution in the glasma. As shown in Fig.~\ref{fig:3dtwopartcorr}, the initial correlation is a peak $\mathcal{C}(\tau_\mathrm{form})\propto\delta(\Delta\phi-\pi)\delta(\Delta\eta)$, which quickly starts to decorrelate in $\Delta\eta$ at $\Delta\tau=0.01\,\mathrm{fm/c}$, followed by a decorrelation in $\Delta\phi$ at $\Delta\tau=0.1\,\mathrm{fm/c}$. We do not propose this as a realistic model but find it suitable for investigating rapidity broadening. Moreover, since the glasma is boost-invariant, any $\eta$-dependent initial model would give the same broadening of the $\Delta\eta$ distribution compared to the initial $\eta$ distribution. Around the typical switching time to a kinetic theory followed by a hydrodynamics description \cite{Kurkela:2018vqr} at $\Delta\tau=0.2-0.3\,\mathrm{fm/c}$, most of the initial correlation gets significantly reduced. The results represented in Fig.~\ref{fig:3dtwopartcorr} correspond to charm quarks with $p_T(\tau_\mathrm{form})=1\,\mathrm{GeV}$ as they propagate in a glasma characterized by $Q_s=2\,\mathrm{GeV}$. 

\begin{figure*}[!hbt]
\includegraphics[width=0.8\textwidth]{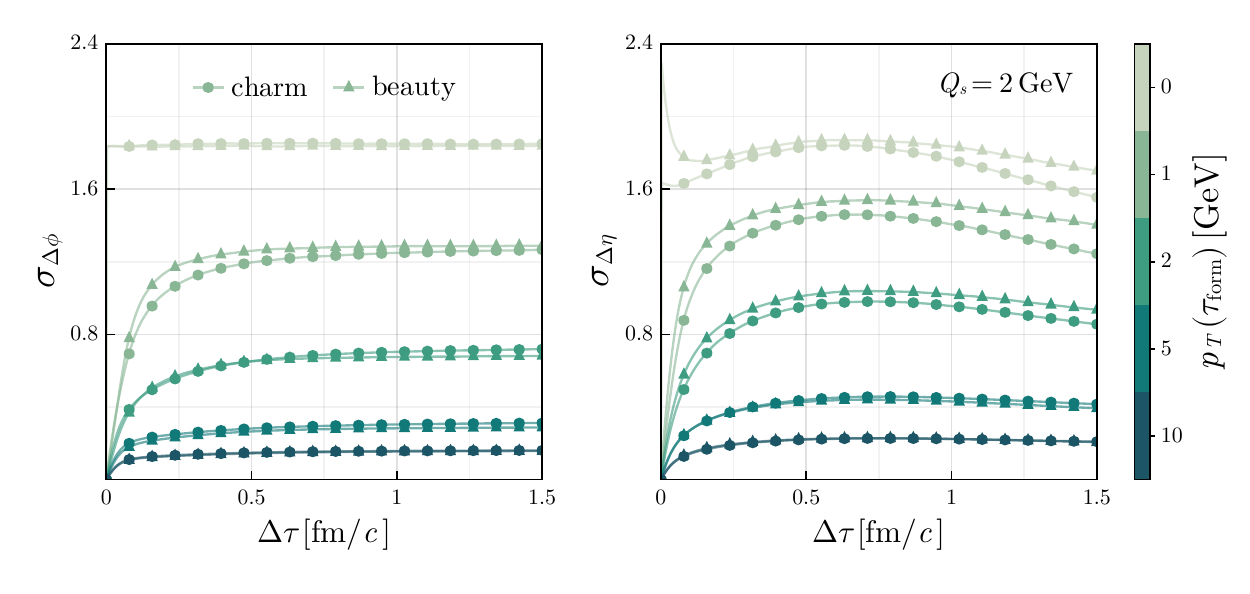}
\caption{\label{fig:sigma_dphideta_tau_pT_dep} Azimuthal $\sigma_{\Delta\phi}$ and rapidity $\sigma_{\Delta\eta}$ decorrelation widths, as a function of proper time $\Delta\tau$, for charm \textit{(circle markers)} and beauty \textit{(triangle markers)}, initialized with $p_T(\tau_{\mathrm{form}})\in\{0, 1, 2, 5, 10\}\,\gev$ \textit{(different colors)}. In all these simulations, the glasma background field has $Q_s=2\,\gev$.}
\end{figure*}

We first show in Fig.~\ref{fig:dndphi_tau}  the azimuthal angle decorrelation of charm and beauty quark pairs initialized with $p_T=2\,\mathrm{GeV}$ in the glasma. More concisely, we numerically extract $\mathcal{C}(\Delta\phi)$ as a function of $\Delta\phi$ at various values of the proper time $\Delta\tau$. The results represented in Fig.~\ref{fig:dndphi_tau} show that the initial correlation $\mathcal{C}(\Delta\phi)\propto \delta(\Delta\phi-\pi)$ rapidly decreases until $\Delta\tau=0.05\,\mathrm{fm/c}$, when the glasma fields are strong. This decrease becomes slower for $\Delta\tau>0.1\,\mathrm{fm/c}$ characterized by dilute fields. The rapidity correlation $\mathcal{C}(\Delta\eta)$ shows similar behaviour. 

To systematically quantify the decorrelation in the glasma and to study the initial particle $p_T(\tau_\mathrm{form})$ and glasma $Q_s$ dependence of the correlation, we extract the rapidity $\sigma_{\Delta\eta}$ and azimuthal $\sigma_{\Delta\phi}$ decorrelation widths, as a function of the time passed from the formation of the quark $\Delta\tau$. The decorrelation widths are extracted from Eq.~\eqref{eq:cdetacdphi} and the procedure is depicted in Fig.~\ref{fig:sketch_decorrelation}. The dependence of the decorrelation widths $\sigma_{\Delta\phi}$ and $\sigma_{\Delta\eta}$ on the quark $p_T$ at formation time $\tau_\mathrm{form}$ is shown in Fig.~\ref{fig:sigma_dphideta_tau_pT_dep} as a function of the relative proper time $\Delta\tau$, for both charm and beauty quarks. The longitudinal decorrelation $\sigma_{\Delta\eta}$ in rapidity is systematically greater than the azimuthal $\sigma_{\Delta\eta}$, which is extracted from the transverse momenta, as expressed in Eq.~\eqref{eq:deltaphi}. Since the glasma initially consists of only longitudinal color fields, quantities affected by the longitudinal dynamics tend to be larger than the corresponding transverse ones. We reported a similar ordering in our previous work \cite{Avramescu:2023qvv} for momentum broadening components $\langle \delta p^2_L\rangle>\langle \delta p^2_T\rangle$ and we interpret $\sigma_{\Delta\eta}>\sigma_{\Delta\phi}$ as arising from the same anisotropic glasma dynamics. As expected, the widths $\sigma_{\Delta\eta},$  $\sigma_{\Delta\phi}$ decrease with increasing $p_T$.
This is natural since heavy quark pairs initialized with small $p_T$ only require a small momentum transfer from the glasma fields to  decorrelate, while pairs with sufficiently large $p_T$ still preserve the initial correlation throughout the evolution. In realistic collisions, assuming heavy quarks produced according to the FONLL distribution from Eq.~\eqref{eq:fonllsigma}, most quarks have small initial $p_T$ and thus one expects the glasma stage to completely decorrelate the bulk of $Q\overline{Q}$ pairs. However, it is experimentally possible to also access the angular correlation between particles at intermediate and high $p_T$ values that do not completely lose their correlation, hence could carry novel and relevant information about the heavy quark in medium diffusion and energy loss.

\begin{figure*}[!hbt]
\includegraphics[width=0.8\textwidth]{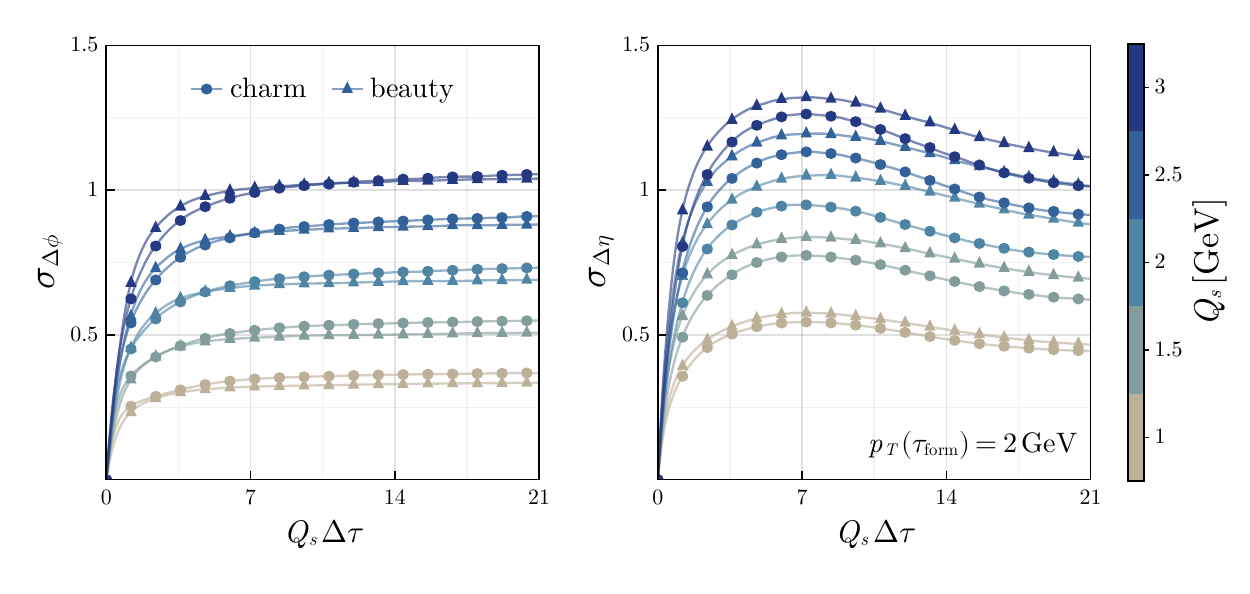}
\caption{\label{fig:sigma_dphideta_tau_Qs_dep} Dependence of the decorrelation width in relative azimuthal angle $\sigma_{\Delta\phi}$ and rapidity $\sigma_{\Delta\eta}$ on the scaled dimensionless proper time $Q_s\Delta\tau$, for charm \textit{(circle markers)} and beauty \textit{(triangle markers)} quarks. The heavy quarks are initialized with the same $p_T=2\,\gev$ but the glasma saturation momentum is varied $Q_s\in\{1, 1.5, 2, 2.5, 3\}\,\gev$ \textit{(different colors)}.}
\end{figure*}

Figure~\ref{fig:sigma_dphideta_tau_Qs_dep} shows a similar analysis for the $Q_s$ dependence of $\sigma_{\Delta\phi}$ and $\sigma_{\Delta\eta}$, while keeping the initial momenta fixed to the moderate value $p_T(\tau_\mathrm{form})=2\,\mathrm{GeV}$, for both charm and beauty quarks. The decorrelation widths in Fig.~\ref{fig:sigma_dphideta_tau_Qs_dep} are shown in terms of $Q_s\Delta\tau$, the dimensionless relevant temporal scale in the glasma. Although in general the saturation time of the glasma dynamics is considered to be $1/Q_s$, our quantitative study in SU(3) shows that the saturation of both the azimuthal and longitudinal widths occurs at $\tau \simeq 3/Q_s$. In the case of typical $Q_s$ values for AA collisions, this implies $\tau\simeq 0.3$ fm/c, which is in good agreement with the typical initial time of the hydrodynamic and transport simulation of the QGP dynamics \cite{Heinz:2013th,Ruggieri:2013ova,Plumari:2015cfa}. 

The anisotropic ordering $\sigma_{\Delta\eta}>\sigma_{\Delta\phi}$ is preserved irrespective of $Q_s$, while the peak in $\sigma_{\Delta\eta}$ gets more pronounced with increasing $Q_s$. Both decorrelation increase with $Q_s$. Since the saturation momentum of the glasma directly dictates the strength of the color fields, one expects more decorrelation of the $Q\overline{Q}$ pairs with larger $Q_s$, as confirmed by our results.

The effect of the decorrelation in Figs. 7 and 8 is larger for beauty than for charm quarks due to the choice $\tau_\mathrm{form}\sim 1/m$. The ordering $m_b>m_c$ would naively suggest that the beauty quarks experience fewer momentum kicks from the glasma fields, resulting in a reduced net decorrelation. Nevertheless, $\tau_\mathrm{form}^b<\tau_\mathrm{form}^c$ thus the beauty quarks experience much stronger glasma fields. The fine interplay between these effects in the end causes beauty quark pairs to decorrelate more than charm quarks. We emphasize that a similar effect was noticed for the larger momentum broadening of beauty quarks in the glasma as compared to charm quarks in \cite{Avramescu:2023qvv}.

We simplify our setup by assuming standard pQCD for heavy quark production with a heuristic $\tau_\mathrm{form}$ formation time, followed by the evolution in the glasma fields. The heavy quark production in pQCD is based on the factorization between hard scattering and the medium. The quark is assumed to be free and on-shell immediately after production. The heuristic $\tau_\mathrm{form}\sim 1/m_c$, is used to model the timescale over which the quark transitions from a virtual particle to a propagating on-shell state. On the other hand, the glasma fields decorrelate on a timescale $\tau_\mathrm{corr}\sim 1/Q_s$, comparable to the charm quark's $\tau_\mathrm{form}\sim 1/m_c$ when $m_c\sim Q_s$. Moreover, the glasma fields evolve rapidly, and the quark may interact with these fields even during its formation phase. These could challenge the assumed separation between the production and in-medium propagation of the heavy quarks, especially for charm quarks. Nevertheless, the azimuthal decorrelation widths for charm and beauty quarks in Fig. 8 show no significant quark mass and thus formation time effect. We conclude that for this observable, assuming a pQCD formation with a heuristic formation time in glasma provides reasonable results for $c\overline{c}$ and $b\overline{b}$ correlations. In the future, an improvement could be provided by the coupling between the quark production in the strong glasma fields and solving the Dirac equation in these fields.

The observed effect of the glasma on the azimuthal decorrelation is similar for beauty quarks, as inferred from Figures 7 and 8. However, beauty quarks are not in kinetic equilibrium with the bulk, given their much longer thermalization timescale (approximately $m_b/m_c$ times larger than that of charm quarks). Thus, for beauty quarks equilibrium effects could not fully wash out the observed correlation in the glasma phase. It is conceivable that correlations could be reduced at $p_T=2\,\mathrm{GeV}$, particularly given that initial production is not perfectly back-to-back in a realistic scenario. Nevertheless, the situation differs at higher transverse momentum, for instance at $p_T=4\,\mathrm{GeV}$. At such momenta, the significant back-to-back correlations are expected to persist, as heavy-flavor quarks are far from full kinetic equilibrium.

\subsection{Nuclear modification factor}

\subsubsection{Numerical $R_{AA}$ (glasma simulations)}

Using the procedures presented in Sec.~\ref{subsec:raa}, we simulate ensembles of heavy quark test particles initialized with a $p_T$ distribution according to the FONLL calculation from Eq.~\eqref{eq:fonllsigma} and follow their evolution in the glasma. As previously done in the literature \cite{Ruggieri:2018rzi,Liu:2019lac,Sun:2019fud}, but now using SU($3$) as a gauge group, we extract the nuclear modification factor in the glasma produced in an $AA$ collision by initializing heavy quarks according to a $pp$ input FONLL calculation, that is we numerically compute $\mathrm{d}N/\mathrm{d}p_T(\tau; pp)$ from Eq.~\eqref{eq:raa}. Additionally, we improve this choice by using an initial FONLL spectrum in $AA$, which contains nuclear PDF effects, to initialize heavy quark production in the glasma, namely $\mathrm{d}N/\mathrm{d}p_T(\tau; AA)$ in Eq.~\eqref{eq:raa}. Unless otherwise specified, we extract $R_{AA}$ using the former initialization with FONLL in $pp$ in order to focus on the effect of the glasma phase. The default choice for the saturation momentum in glasma is $Q_s=2\,\mathrm{GeV}$, the input FONLL calculation in $pp$ corresponds to $\sqrt{s_\mathrm{pp}}=5.5\,\mathrm{TeV}$ and the PDF choice CTEQ6.6. 

\begin{figure*}[!hbt]
\includegraphics[width=0.8\textwidth]{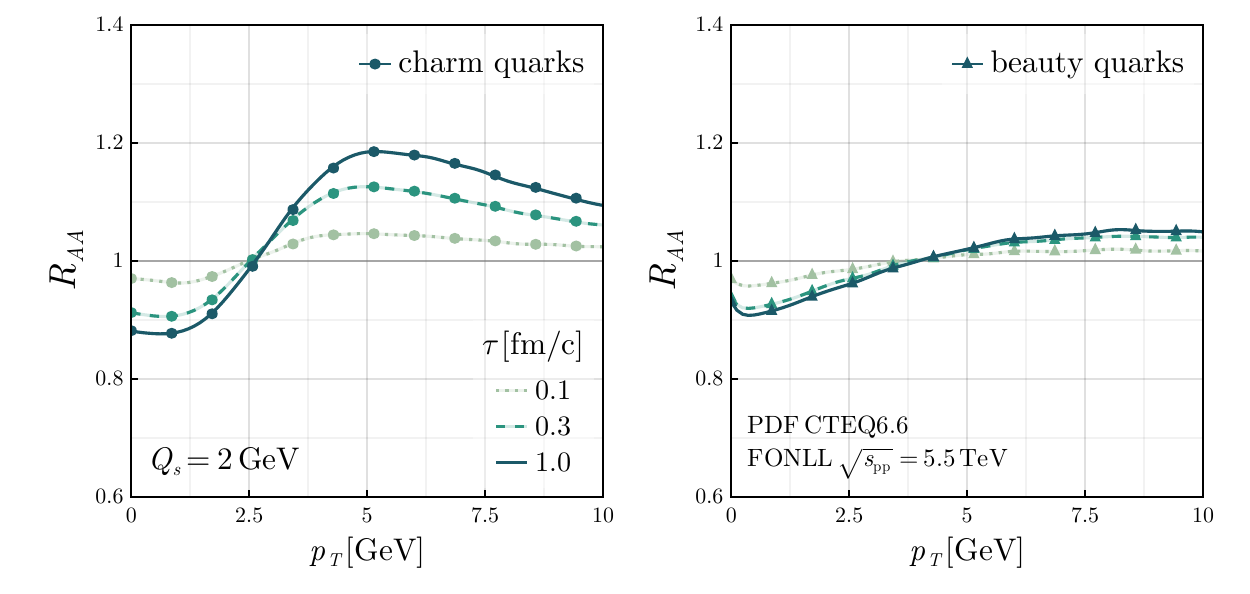}
\caption{\label{fig:raa_tau_dep_charm_beauty} Nuclear modification factor $R_{AA}$ computed using Eq.~\eqref{eq:raa}, as a function of $p_T$, at various $\tau$ values (\textit{different colors and line styles}). The results are shown for both charm (\textit{circle markers}) and beauty (\textit{triangle markers}) evolving in a glasma with $Q_s=2\,\mathrm{GeV}$ and produced according to the FONLL distribution from Eq.~\eqref{eq:fonllsigma} using the PDF set CTEQ6.6 at the collision energy $\sqrt{s_\mathrm{pp}}=5.5\,\mathrm{TeV}$.}
\end{figure*}

The temporal evolution of $R_{AA}(p_T)$ is shown in Fig.~\ref{fig:raa_tau_dep_charm_beauty} for both charm and beauty quarks, at various proper time regimes, namely the very-early stage $\tau=0.1\,\mathrm{fm/c}$, the typical switching time to the subsequent stage $\tau=0.3\,\mathrm{fm/c}$ and the late stage $\tau=1\,\mathrm{fm/c}$. The profile of $R_{AA}$ is dictated by $p_T$ migration from low to high $p_T$ values, as revealed by the Gaussian broadening from the toy model initialization with fixed $p_T$, see Fig.~\ref{fig:sketch_raa_gl_fonll}. This causes a depletion of particles in the small-$p_T$ range of about $p_T<2.5\,\mathrm{GeV}$ for charm and $p_T<4\,\mathrm{GeV}$ for beauty quarks. Since the total number of quarks is conserved, the $p_T$ migration causes an enhancement of $R_{AA}$ at larger $p_T$. For beauty quarks, at sufficiently large $p_T$ values, $R_{AA}$ reaches a plateau as an effect of $p_T$ migration equalization from larger and smaller $p_T$ values, but for charm quarks $R_{AA}$ starts to decrease at large $p_T$ for later times $\tau$. 

At later times $\tau$, the \mbox{small-$p_T$} migration followed by a large-$p_T$ enhancement becomes more pronounced since the test particles get increasingly more deflected by the color force of the glasma fields. As previously observed in \cite{Avramescu:2023qvv}, since beauty quarks are formed earlier and thus experience stronger glasma fields, their $p_T$ momentum broadening is larger than for charm quarks. Nevertheless, the outcome of our simulations is that $R_{AA}$ is closer to identity for beauty quarks. This effect may be traced back to the $p_T$ spectrum of charm quarks being steeper at large $p_T$, resulting in more $p_T$ migration and consequently an enhanced $R_{AA}$ as compared to beauty quarks. 

\begin{figure*}[!hbt]
\includegraphics[width=0.8\textwidth]{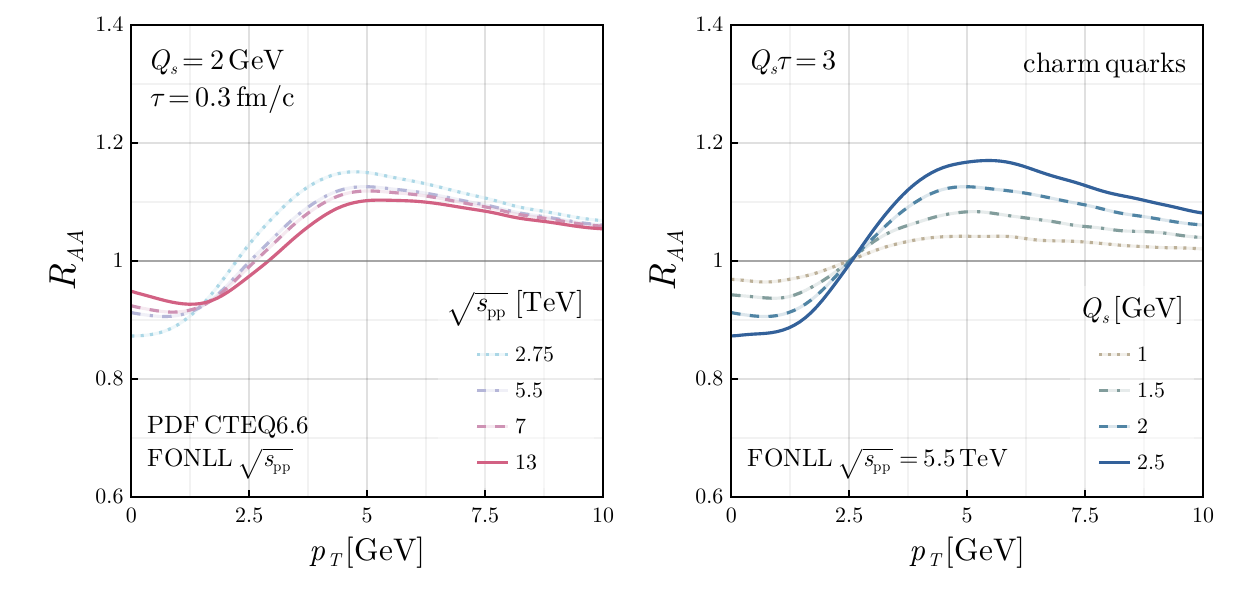}
\caption{\label{fig:raa_charm_Qs_energy_dep} Initial FONLL energy $\sqrt{s_\mathrm{pp}}$ (\textit{left panel}) and glasma saturation momentum $Q_s$ (\textit{right panel}) dependence of the nuclear modification factor $R_{AA}(p_T)$ for charm quarks. In the (\textit{left panel}) $R_{AA}$ is extracted at $\tau=0.3\,\mathrm{fm/c}$ with fixed glasma $Q_s=2\,\mathrm{GeV}$ and varying FONLL energy $\sqrt{s_\mathrm{pp}}$ (\textit{different colors and line styles}) for the PDF set CTEQ6.6, while in the (\textit{right panel}) $Q_s$ is varied (\textit{different colors and line styles}) and $R_{AA}$ is represented at fixed $Q_s\tau$ and FONLL energy $\sqrt{s_\mathrm{pp}}=5.5\,\mathrm{TeV}$ and the same PDF set.}
\end{figure*}

The $\sqrt{s_\mathrm{pp}}$ dependence of $R_{AA}$ results from a combination of two separate effects. The collision energy $\sqrt{s_\mathrm{pp}}$ influences the initial heavy quark spectrum in the FONLL calculation. On the other hand, also the saturation scale $Q_s$ depends on $\sqrt{s_\mathrm{pp}}$. We will first quantify these two effects separately, studying 
the dependence of $R_{AA}$ on the initial $\sqrt{s_\mathrm{pp}}\in\{2.75, 5.5, 7, 13\}\,\mathrm{TeV}$ for fixed $Q_s=2\,\mathrm{GeV}$ and on $Q_s\in\{1, 1.5, 2, 2.5\}\,\mathrm{GeV}$ using a given $\sqrt{s_\mathrm{pp}}=5.5\,\mathrm{TeV}$ in the FONLL calculation, all with the same PDF set. The results depicted in Fig.~\ref{fig:raa_charm_Qs_energy_dep} show a stronger dependence of $R_{AA}$ on $Q_s$ than on $\sqrt{s_\mathrm{pp}}$. The FONLL spectra become less steep in the large-$p_T$ tail with increasing collision energy $\sqrt{s_\mathrm{pp}}$, yielding a smaller effect on the magnitude of $R_{AA}$. Increasing the saturation momentum $Q_s$ produces stronger color fields, and thus generates larger $p_T$ broadening and more enhancement in $R_{AA}$. 

\begin{figure}[!hbt]
\includegraphics[width=0.8\columnwidth]{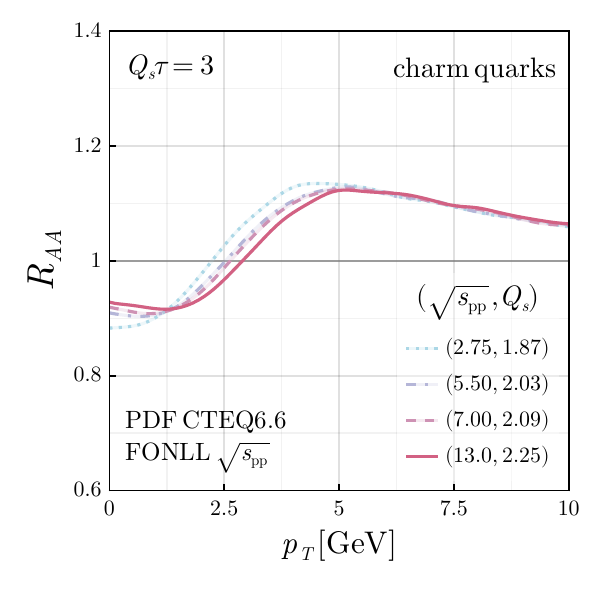}
\caption{\label{fig:raa_charm_Qs_energy_dep_map} Initial FONLL energy $\sqrt{s_\mathrm{pp}}$ and saturation momentum $Q_s$ dependence (\textit{different colors and line styles}) of $R_{AA}(p_T)$, using the mapping from Eq.~\eqref{eq:smapstoQs} for charm quarks at $Q_s\tau=3$, using the PDF set CTEQ6.6.}
\end{figure}

To combine the two sources of $\sqrt{s}$-dependence of $R_{AA}$, we need to explicitly specify the dependence of the gluon saturation momentum $Q_s$ on the collision energy $\sqrt{s}$. For this purpose, we use a mapping $\sqrt{s}\mapsto Q_s$ proposed in~\cite{Lappi:2007ku}. The nuclear saturation momentum scales according to a nuclear geometry factor with respect to the proton one $Q^2_{s,A}\propto g(A)Q^2_{s,p}$, in which we choose $g(A)\propto A^{1/3}$ and neglect $\log A$ contributions. Within the GBW parametrization of the deep inelastic scattering (DIS) cross section \cite{Golec-Biernat:1998zce}, the proton saturation momentum may be parametrized in terms of the $x$ momentum fraction of the parton as $Q_{s,p}^2=Q_0^2(x_0/x)^\lambda$. The values of the involved parameters are obtained from a fit to HERA data as $\lambda=0.277$, $x_0=0.41\cdot 10^{-4}$ and $Q_0=1\,\mathrm{GeV}$~\cite{Golec-Biernat:1998zce}. In practice, one chooses an effective $x$ value for gluons as $x= x_\mathrm{eff}$ where $x_\mathrm{eff}\sim Q_s/\sqrt{s}$. Collecting all these parameterizations yields a mapping for the nuclear saturation momentum in terms of the collision energy which is valid up to a proportionality factor that may not be determined from our current setup. In this work, we choose to parametrize the saturation momentum as
\begin{equation}
    \label{eq:smapstoQs}
    Q^2_{s,A}=A^{1/3}(Q_0^2\,x_0^\lambda\,\sqrt{s}^\lambda)^{2/(2+\lambda)}.
\end{equation}
We use this mapping and obtain the results shown in Fig.~\ref{fig:raa_charm_Qs_energy_dep_map}. Here one sees that the $\sqrt{s}$-dependences of the FONLL spectrum and $Q_s$ partially cancel, resulting in a a weak overall $\sqrt{s}$-dependence. In particular, the combined effect leads to very little variation of the magnitude of $R_{AA}$ with $\sqrt{s}$, but to shift towards higher $p_T$.
Comparing with the independent variations of $Q_s$ and $\sqrt{s}$ in the FONLL spectrum, one may infer that the  $\sqrt{s}$-dependence in FONLL is a slightly stronger effect.

\begin{figure}[!hbt]
\includegraphics[width=0.8\columnwidth]{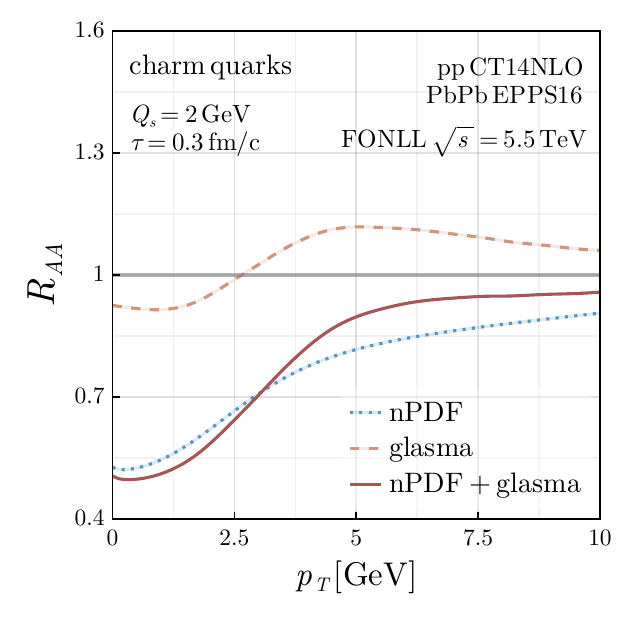}
\caption{\label{fig:raa_charm_pdf_vs_npdf} Nuclear modification factor $R_{AA}(p_T)$ extracted from Eq.~\eqref{eq:raa} using either the $pp$ FONLL calculation with CT14NLO PDF as glasma input, labeled as \textit{glasma}, or the $AA$ FONLL calculation with EPPS16 nPDF in the glasma particle initialization, denoted as \textit{nPDF+glasma}. The effect arising solely from nuclear effects in the FONLL calculation is also represented, with the label \textit{nPDF}. The calculations are done for charm quarks evolving in a glasma with $Q_s=2\,\mathrm{GeV}$, in a collision of $\sqrt{s}=5.5\,\mathrm{TeV}$, at the proper time $\tau=0.3\,\mathrm{fm/c}$.} 
\end{figure}

We can also account for nuclear PDF effects by initializing the heavy quark production probability density in the glasma $\mathrm{d}N/\mathrm{d}p_T(\tau;AA)$ with the FONLL calculation from Eqs.~\eqref{eq:fonllsigma} and~\eqref{eq:fonlldndpt} for $AA$ collisions. The results for $R_{AA}$ extracted using Eq.~\eqref{eq:raa} for the combined nPDF + glasma effect are depicted in Fig.~\ref{fig:raa_charm_pdf_vs_npdf} and compared with the previous glasma extraction of $R_{AA}$ using the glasma spectrum initialized with FONLL in $pp$, $\mathrm{d}N/\mathrm{d}p_T(\tau;pp)$, and with $R_{AA}$ resulting solely from nuclear PDFs. The key finding is that the glasma $p_T$ broadening enhances the nuclear effect. In both calculations including nPDFs, $R_{AA}<1$ due to the nuclear shadowing phenomenon relevant in the small-$x$ regime \cite{Mueller:1985wy,Eskola:2009uj,Eskola:2016oht}. Since the overall effect of the glasma fields is to induce $p_T$-migration from low to high values, the $p_T$-differential slope of $R_{AA}$ is larger for the nPDF + glasma result compared to just the nPDF calculation. 
It was noted in \cite{Sun:2019fud} that glasma dynamics (already in SU(2)) would induce a shape of $R_{AA}(p_T)$ that increases with $p_T$ which is opposite to what one would get considering heavy quark scatterings in the QGP already from $\tau=0^+$. Here we have found that nPDF+glasma in SU(3) generate globally quite a stronger modification of $R_{AA}(p_T)$ than the one considered in \cite{Sun:2019fud}. This can be expected to significantly affect the quantitative estimates of the heavy quark space diffusion coefficient $D_s(T)$ and its relation to observables like $R_{AA}(p_T)$ and $v_2(p_T)$.

\subsubsection{Analytical $R_{AA}$ (toy model)}

\begin{figure}[!hbt]
\includegraphics[width=0.8\columnwidth]{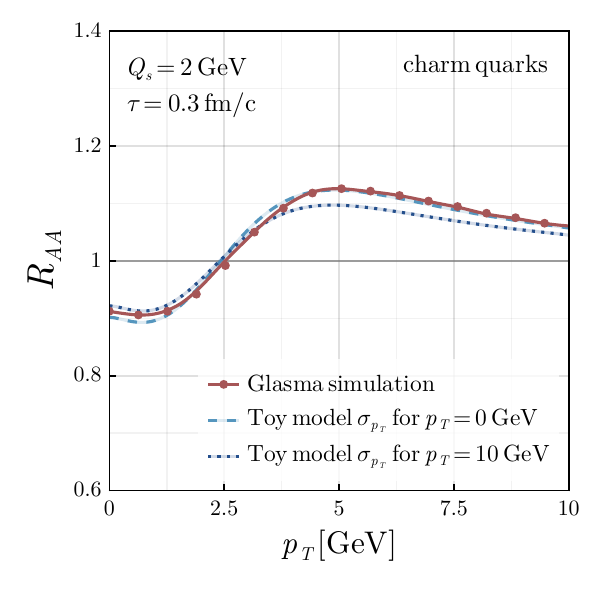}
\caption{\label{fig:raa_charm_toy_vs_glasma} Nuclear modification factor $R_{AA}(p_T)$ extracted from glasma simulations (\textit{circle markers}) or from the analytical toy model using a Gaussian momentum exchange with fixed widths $\sigma_{p_T}$ \textit{(different styles of lines)}. Results are shown for charm quarks at $\tau=0.3\,\mathrm{fm/c}$ in a glasma with $Q_s=2\,\mathrm{GeV}$ and a FONLL calculation with $\sqrt{s_\mathrm{pp}}=5.5\,\mathrm{TeV}$ with the CTEQ6.6 PDF set.}
\end{figure}

The Gaussian $\vec{p}_T$ broadening toy model described in Sec.~\ref{subsec:analtoyraa} allows the analytical extraction of $R_{AA}$ defined in Eq.~\eqref{eq:raatoy} using the glasma probability density proposed in Eq.~\eqref{eq:dndptconv}. The only input required for this calculation is the width of the Gaussian broadening. For this purpose, we approximate $\sigma$ with the standard deviation of the momentum kicks $\delta p_T^2$ expressed in Eq.~\eqref{eq:mombroad}. We denote this width as $\sigma_{p_T}$ and we extract it from numerical glasma simulations. For simplicity, we perform this extraction using charm quarks initialized with $p_T(\tau_\mathrm{form})=0\,\mathrm{GeV}$ which gives $\sigma_{p_T}=0.64\,\mathrm{GeV}$ and $p_T(\tau_\mathrm{form})=10\,\mathrm{GeV}$ for which $\sigma_{p_T}=0.57\,\mathrm{GeV}$. The results for the analytical toy model $R_{AA}$ are shown in Fig.~\ref{fig:raa_charm_toy_vs_glasma} and compared with the numerical glasma $R_{AA}$. The shape of the $p_T$ dependence is well reproduced while the magnitude slightly varies depending on the value of the Gaussian width $\sigma_{p_T}$. This result shows that, for reproducing $R_{AA}$, the Gaussian $p_T$ broadening toy model captures the dynamics in the glasma phase.

\section{Conclusion}
\label{sec:conclusion}

We studied the phenomenological impact of the early-stage glasma fields on heavy quark angular correlations and transverse momentum spectra. Using our numerical solver for the transport of classical test particles in the non-Abelian Yang-Mills fields of the glasma, developed and tested in our previous work \cite{Avramescu:2023qvv}, we simulate ensembles of heavy quarks evolving in the glasma background fields. This study contains the first extraction of heavy quark two-particle correlations in the pre-equilibrium stage and a refined calculation of heavy quark $R_{AA}$ including both glasma and nPDF effects. 

First, we extract the two-particle correlations of heavy quark $Q$ and antiquark $\overline{Q}$ pairs initially produced back-to-back in the glasma. We quantify how the glasma changes the initial correlation by extracting the rapidity $\sigma_{\Delta\eta}$ and azimuthal $\sigma_{\Delta\phi}$ correlation widths at different proper times $\Delta\tau$. Further, we systematically study the dependence of the decorrelation widths on the initial heavy quark $p_T$ and glasma saturation momentum $Q_s$. Our results show that the glasma stage has a large effect on the two-particle correlations of low to moderate-$p_T$ heavy quarks, causing a rapid decorrelation in both rapidity $\Delta\eta$ and relative azimuthal angle $\Delta\phi$. This effect gets more pronounced with increasing saturation momentum $Q_s$, since it encodes the strength of the glasma fields and thus the magnitude of the momentum kicks experienced by the heavy quarks. Our main finding is that the azimuthal correlation of $c\overline{c}$ pairs is significantly modified by the glasma phase. Since this result is only applicable in the early stage, 
combining our results with a modeling of the subsequent stages will be highly relevant for comparisons with experimental measurements. For example, the azimuthal correlations of $D\overline{D}$ pairs will be measured in heavy ion collisions at ALICE3~\cite{ALICE:2022wwr} to directly probe momentum broadening in the QGP, and azimuthal correlations of muons originating from heavy-flavor decays in PbPb collisions  were reported by ATLAS \cite{ATLAS:2023vms}. 

Second, we investigated the effect of the glasma stage on the heavy quark nuclear modification factor $R_{AA}$. Earlier studies \cite{Ruggieri:2018rzi,Liu:2019lac,Sun:2019fud} evaluate either $R_{AA}$ or $R_{pA}$ in the glasma by initializing heavy quarks according to a $p_T$ distribution given by the pQCD FONLL calculation in $pp$ and then extract the resulting $\mathrm{d}N/\mathrm{d}p_T$ in an SU($2$) glasma. We improve these calculations by using SU($3$) as a gauge group, initializing the heavy quark spectra with the FONLL calculation in both $pp$ and $AA$. Our results qualitatively confirm the findings of these earlier works, namely that the glasma phase induces an overall $p_T$ migration from small to large $p_T$ for heavy quarks, causing $R_{AA}<1$ at smaller-$p_T$ and $R_{AA}>1$ at larger-$p_T$, to conserve the total number of heavy quarks. Further, we study the dependence of $R_{AA}$ on the saturation momentum $Q_s$ and the initial collision energy $\sqrt{s}$ in the FONLL calculation, along with a more realistic mapping between $Q_s$ and $\sqrt{s}$ inspired by DIS, and observe a large cancellation between the $\sqrt{s}$-dependences of the FONLL spectrum and the value of $Q_s$. Additionally, we compare the numerically extracted $R_{AA}$ with a toy model which assumes symmetric Gaussian $\vec{p}_T$ broadenings. We find that such an approximation is only qualitatively in agreement with the results of the numerical simulations and can not replace the full glasma simulation. Lastly, we improve our framework by incorporating nPDF effects in the glasma $R_{AA}$. We achieve this by initializing heavy quark $p_T$ spectra according to the FONLL cross-section in $AA$ collisions, a calculation which includes nuclear PDF effects. Our results yield a combined nPDF+glasma $R_{AA}<1$ caused by nuclear shadowing, and the nuclear modification is found to be more pronounced compared to the nPDF-only setup.  This calculation for the heavy quark $R_{AA}$ contains the most important initial stage effects, arising from CGC-inspired early-stage dynamics and nPDF considerations. 

Our extraction of the two-particle correlations relies on simple glasma field configurations and toy model heavy quark test particle initialization. We plan to extend our analysis by including more sophisticated field and particle initial conditions. For the glasma fields produced in $AA$ collisions, more realistic geometries may be achieved by incorporating finite-size effects. For this purpose, one may use an impact parameter $b$ dependent MV model \cite{Schenke:2012wb,Schlichting:2014ipa,Mantysaari:2017cni,Mantysaari:2022sux}, where the initial MV color charge is sampled according to a Woods-Saxon impact parameter $b$-dependent distribution. This yields glasma fields whose energy density is contained within a finite simulation region. Additionally, we are interested in extending the $Q\overline{Q}$ angular decorrelation study to small systems, for which there is experimental data on $D\overline{D}$ angular correlations in either $pp$ \cite{CMS:2011yuk,LHCb:2012aiv,LHCb:2017bvf,ATLAS:2017wfq,ALICE:2019oyn} or $pA$ \cite{LHCb:2020jse,ALICE:2019oyn}. For this purpose, we plan to incorporate proton geometry \cite{Schlichting:2014ipa,Mantysaari:2016ykx,Mantysaari:2020axf,Kumar:2022aly} in our glasma implementation. With this setup, one could further study the effect on the two-particle correlations of nucleonic and sub-nucleonic fluctuations or hot spots in the high-energy proton \cite{Cepila:2017nef,Mantysaari:2020axf,Demirci:2023ejg,Mantysaari:2022sux}. 

The toy model initialization for the heavy quark coordinate and momenta may also be improved. 
A more realistic coordinate distribution can also be implemented by producing the quarks according to the initial energy density profile of the glasma fields. This will become more important when considering smaller systems like $pp$ or $pA$ collisions and finer geometry details like fluctuations or hot spots. Additionally, one could use a heavy-ion event generator such as Pythia Angantyr~\cite{Bierlich:2018xfw} to input the $p_T$ and angular $\Delta\phi$ distributions of heavy quark $Q\overline{Q}$ pairs in the evolving glasma fields produced in $AA$ collisions. Lastly, one additional effect arising from finite-size effects in the glasma is that the $Q\overline{Q}$ pairs may experience a different decorrelation since at least one of the quarks from the pair could in principle ``escape'' the highly dense medium. This will require a more sophisticated modeling of the glasma transverse density profile which will produce a different angular decorrelation for $Q\overline{Q}$ pairs evolving in this profile. Such an implementation would be of interest for further studies. 

One may go beyond the boost-invariant approximation implemented at the level of the glasma fields employed in the current study, by evolving the glasma fields in a full 3+1D setup \cite{Schenke:2016ksl,Gelfand:2016yho,Ipp:2017lho,Schlichting:2020wrv,Ipp:2020igo,Ipp:2021lwz,McDonald:2023qwc,Matsuda:2023gle,Ipp:2024ykh,Matsuda:2024moa}. Such an improvement will enable the study of the 3D angular decorrelation of the $Q\overline{Q}$ pairs in a 3+1D background field and will provide more reliable results for the rapidity dependence of $\mathcal{C}(\Delta\eta)$. Moreover, beyond the extraction of the two-particle correlation, the effect of the rapidity-dependent glasma fields on the transport of both heavy quarks and jets in the glasma represents a promising direction for future research. 

Another shortcoming of the current study is the lack of an energy loss mechanisms for the heavy quarks. The glasma fields induce a significant momentum broadening for both heavy quarks and jets. Such a momentum broadening causes gluon emission leading to radiative energy loss, which is not accounted for in our present formalism. However, it would be highly desirable to incorporate more realistic energy loss mechanisms in our current framework. For this purpose, the calculations developed for classical gluon radiation from a point particle propagating in CGC fields \cite{Kajantie:2019hft,Kajantie:2019nse,Kolbe:2021dhu} may provide additional insight. Alternatively, one may take inspiration from the energy loss formalism recently developed in \cite{Barata:2024xwy} for jets in a glasma background field, based on synchrotron gluon radiation \cite{Shuryak:2002ai,Zakharov:2008uk}. Lastly, energy loss via gluon emissions may be included by using the framework developed for the quantum evolution of a jet in a classical background field \cite{Li:2020uhl,Li:2021zaw,Li:2023jeh}. Such methods will enable us to study the effect of the glasma fields on the energy loss of hard probes and to quantify it by extracting, for example, the nuclear modification factor $R_{AA}$. Additionally, energy loss effects will be crucial for understanding the jet evolution and in-medium radiation in the early-stages. 

Lastly, it would be of great interest to couple these early-time predictions with the subsequent stages of heavy-ion collisions. Earlier model estimates performed using kinetic theory \cite{Boguslavski:2023alu,Boguslavski:2023fdm,Boguslavski:2023waw}, which is applicable after the glasma stage, managed to couple the transport coefficients $\kappa$ for heavy quarks and $\hat{q}$ for jets. Nevertheless, the very-early stage contribution to the azimuthal correlation $\mathcal{C}(\Delta\phi)$ and nuclear modification factor $R_{AA}$ has not been yet accounted for in a systematic manner in both the glasma and QGP phases. In order to compare to finally measured experimental data, model calculations would require simultaneously coupling multiple stages for both the medium and the hard probes: the glasma background fields to kinetic theory or hydrodynamics, the transport of the hard probes in glasma to other transport models applicable during the QGP phase, and the interaction of the hard probes with the underlying medium throughout its evolution. Even though a complete unification of all stages is not yet feasible, our current approach may be extended to provide an improved input to kinetic theory. Coupling the glasma to kinetic theory may be achieved by extracting the gluon distribution function of the glasma fields, as previously done in \cite{Berges:2013fga,Greif:2017bnr}. For the hard probes, including a collision term in the Boltzmann-Vlasov equation used for the classical transport of the probes would provide a more correct treatment of the interactions between the hard degrees of freedom \cite{Dumitru:2005gp,Dumitru:2005hj,Dumitru:2006pz,Dumitru:2007rp,Schenke:2008gg}. 

\begin{acknowledgements}
We are grateful to I.~Helenius and H.~Paukkunen for very helpful discussions about heavy quark production, the FONLL calculation, and nPDFs. D.A.~is grateful to Pol-Bernard Gossiaux for insightful discussions about $D\overline{D}$ correlations. This work was supported by the Research Council of Finland, the Centre of Excellence in Quark Matter (projects 346324 and 364191), and projects 338263, 346567, and 359902 (H.M). D.A.~acknowledges the support of the Vilho, Yrjö and Kalle Väisälä Foundation. V. G. acknowledges support from the European Union – Next Generation EU through the program PRIN2022 (Cod. 2022SM5YAS). D.M.~acknowledges support from the Austrian Science Fund (FWF) projects P~34764 and P~34455. This work was also supported under the European Union’s Horizon 2020 research and innovation programme by the European Research Council (ERC, Grant Agreements  No. ERC-2023-101123801 GlueSatLight and ERC-2018-ADG-835105 YoctoLHC) and by the STRONG-2020 project (Grant Agreement No. 824093). The content of this article does not reflect the official opinion of the European Union and responsibility for the information and views expressed therein lies entirely with the authors.  Computing resources from CSC – IT Center for Science in Espoo, Finland and the Finnish Grid and Cloud Infrastructure (persistent identifier \texttt{urn:nbn:fi:research-infras-2016072533}) were used in this work.
\end{acknowledgements}

\appendix


\section{Classical Casimir invariants} 
\label{appen:casimirs}

\subsection{Casimir scaling}
\label{subappen:casscal}

As expressed in Eq.~\eqref{eq:colorrot}, the heavy quark color charge $Q(\tau)$ is obtained by applying color rotations to an initial color charge vector $Q(\tau_0)\equiv Q_0$. The color components of the initial color charge $Q_0=Q_0^a T^a$ obey the Casimir invariant constraints from Eq.~\eqref{eq:q23}, namely $q_{2}(R)=Q_0^aQ^a_0$ and $q_{3}(R)=d_{abc}Q_0^a Q_0^b Q_0^c$. The values of the quadratic $q_2$ and cubic $q_3$ Casimir invariants are unique to the representation $R$. As discussed in Sec.~\ref{subsec:numparams}, for quarks which lie in the $R=F$ fundamental representation, we choose $q_2=C_2(F)$, with $C_2(F)=(N^2_c-1)/(2N_c)$ the group theory quadratic Casimir in the fundamental representation of SU($N_c$), and $q_3=0$.

There exists an arbitrariness in choosing the values for the classical Casimirs. A natural choice would be to fix their values to the group theory Casimirs. Nevertheless, as noticed in our previous work \cite{Avramescu:2023qvv}, there is no way to construct real classical color charge components $Q^a_0$ whose classical Casimirs correspond to the standard group theory values, namely both $q_{2,3}(R)\neq C_{2,3}(R)$. Nevertheless, it has been observed \cite{Kelly:1994dh, Litim:1999id, Litim:1999ns, Litim:2001db,Avramescu:2023qvv} that for the choice 
$q_{2,3}(R)=D_R\,C_{2,3}(R)$, where $D_R$ is the dimension of the representation, one can sample such classical color charges. 

This choice consequently affects any classical quantities extracted using the classical color charges. Most importantly, using arbitrary classical Casimirs invariants yields classical quantities which do not map to their quantum analogues, extracted using group theory Casimirs. In \cite{Avramescu:2023qvv}, we analyzed such a quantity, more precisely the momentum broadening defined as
\begin{equation}
    \label{eq:mombroad}
    \delta p_i^2(\tau)\equiv p_i^2(\tau)-p_i^2(\tau_\mathrm{form})
\end{equation}
for each component $i=x,y,z$ and then averaged $\langle\delta p^2_i\rangle$ over multiple glasma and particle configurations. We noticed its special scaling property 
\begin{equation}
    \label{eq:casscal}
    \langle \delta p^2_i\rangle_R\propto q_2(R)
\end{equation}
in terms of the classical quadratic Casimir $q_2(R)$. Starting from the initial choice $q_{2,3}=D_R C_{2,3}(R)$, we used this scaling and obtained an expression for extracting the momentum broadening corresponding to group theory Casimirs as $\langle p^2_i\rangle_R/D_R$. Nevertheless, this mapping was derived only for momentum broadening of either infinitely massive or highly energetic quarks and is thus not applicable to other quantities, such as $p_T$-spectra or correlation widths, and heavy quarks beyond the infinite mass approximation.

\subsection{Effect of cubic Casimir} 
\label{subappen:cubiccasimirs}

The Casimir scaling of the momentum broadening from Eq.~\eqref{eq:casscal}, observed in limiting cases, suggests that the cubic Casimir $q_3$ does not affect, or only very weakly affects, the particle momenta. Inspired by this scaling, we investigate whether the classical quantities of interest for us in this study, such as the components of the momentum broadening for quarks in any kinematic regime, the widths of two-particle correlations and the nuclear modification factor, exhibit any dependence on the cubic Casimir $q_3$. A lack of dependence on $q_3$ implies that the effect of the classical color algebra is primarily given by $q_2$. Thus, one may fix $q_2=C_2(F)$ and $q_3=0$, which will yield values compatible with the quantum calculations. To this end, we study how the aforementioned classical quantities depend on the variation of $q_3$. Further, we show only the dependence of the momentum broadening and decorrelation widths on the choice of $q_3$, by keeping $q_2$ fixed with either $q_2=D_F C_2(F)=4$ as done in \cite{Avramescu:2023qvv}, or $q_2= C_2(F)=4/3$ for SU($3$).

\begin{figure*}[!t]
\centering
\subfigure[Scaled momentum broadening $\langle p^2\rangle/D_F$ of colored charges having the quadratic Casimir $q_2=D_F C_2$]{\includegraphics[width=0.75\columnwidth]
{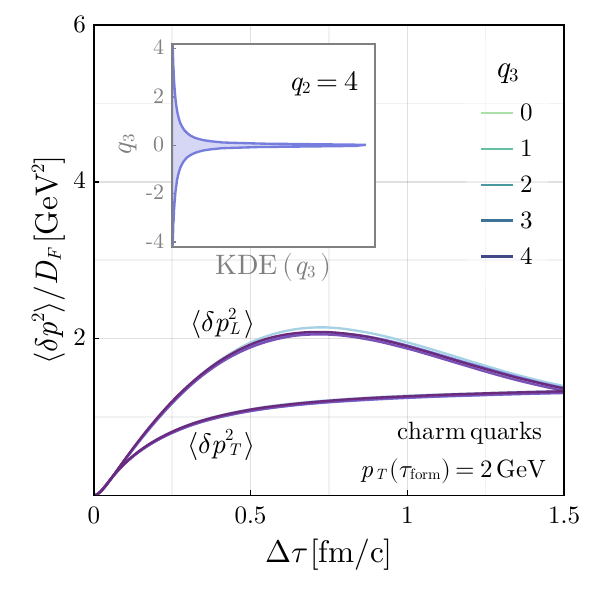}
}
\hspace{0.15\columnwidth}
\subfigure[Momentum broadening $\langle p^2\rangle/D_F$ of colored charges having the quadratic Casimir $q_2=C_2$]{\includegraphics[width=0.75\columnwidth]
{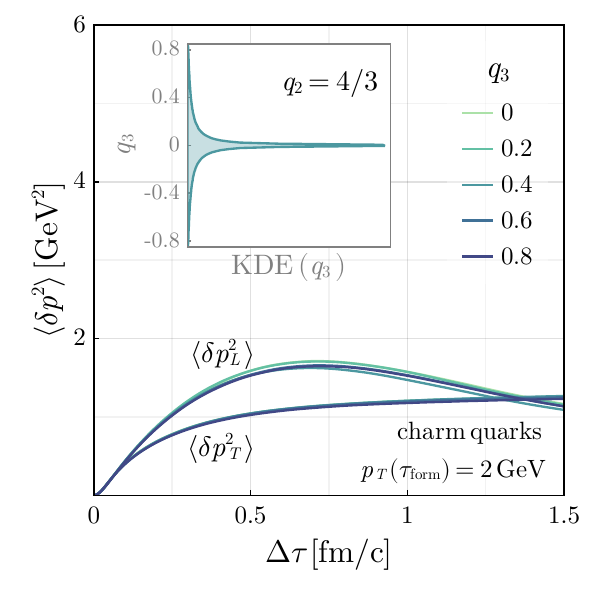}
}
\caption{Proper time $\Delta\tau$ evolution of the longitudinal $\langle\delta p^2_L\rangle$ and transverse $\langle\delta p^2_T\rangle$ components of the momentum broadening for charm quarks with $p_T(\tau_\mathrm{form})=2\,\gev$. The corresponding classical colored charges are initialized with fixed quadratic Casimir $q_2=4$ \textit{(left)} and $q_2=4/3$ \textit{(right)}, while the cubic Casimir $q_3$ is allowed to vary. From all the possible values for $q_3$, as represented in the \texttt{KDE} from the \textit{(insets)}, a few representative ones are chosen to be varied.
}
\label{fig:mombroadq3dep}
\end{figure*}

\begin{figure*}[!t]
\centering
\subfigure[Decorrelation widths $\sigma_{\Delta\phi}$ and $\sigma_{\Delta\eta}$ of colored charges having the quadratic Casimir $q_2=D_F C_2$]{\includegraphics[width=0.75\columnwidth]
{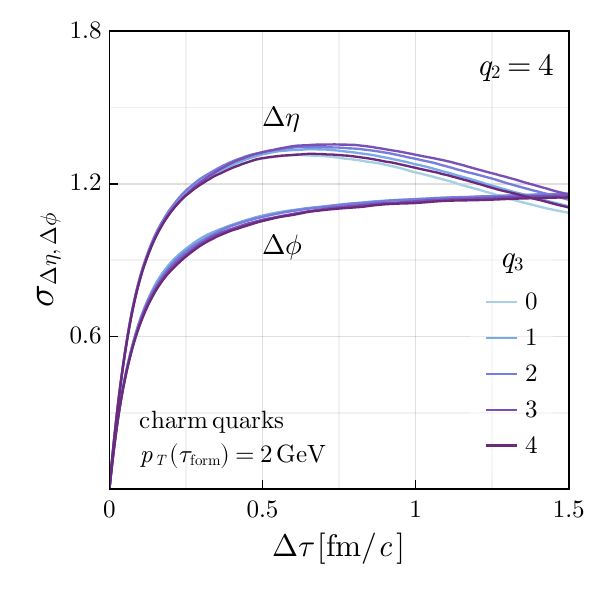}
}
\hspace{0.15\columnwidth}
\subfigure[Decorrelation widths $\sigma_{\Delta\phi}$ and $\sigma_{\Delta\eta}$ of colored charges having the quadratic Casimir $q_2=C_2$]{\includegraphics[width=0.75\columnwidth]
{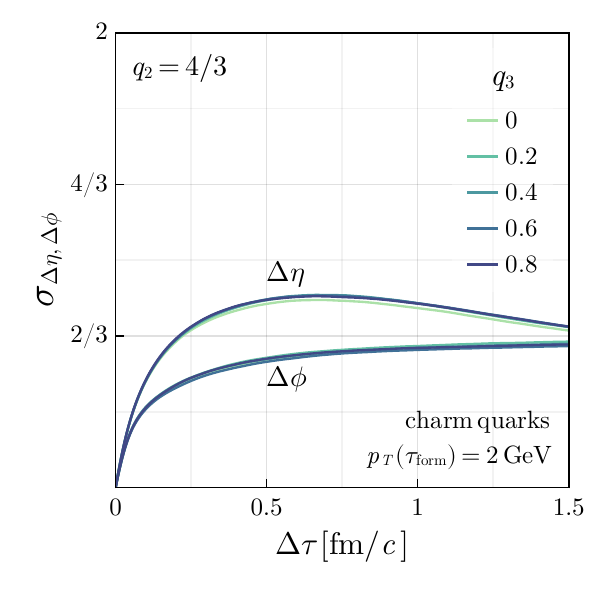}
}
\caption{Proper time $\Delta\tau$ evolution of the decorrelation width in relative azimuthal angle $\sigma_{\Delta\phi}$ and pseudorapidity $\sigma_{\Delta\eta}$ for charm anticharm pairs initialized with $p_T(\tau_\mathrm{form})=2\,\gev$. As in Fig.~\ref{fig:mombroadq3dep}, the cubic Casimir $q_3$ is varied while keeping the quadratic Casimir with the fixed values of $q_2=4$ \textit{(left)} and $q_2=4/3$ \textit{(right)}.}
\label{fig:decorrq3dep}
\end{figure*}

\subsubsection{Momentum broadening}

The dependence on $q_3$ of the longitudinal $L$ and transverse $T$ components of the momentum broadening for charm quarks initialized with $p_T(\tau_\mathrm{form})=2\,\mathrm{GeV}$ is shown in Fig.~\ref{fig:mombroadq3dep}, using either $q_2=4$ or $q_2=4/3$. First, we generate a large ensemble of classical color charge configurations $\left\{Q^a\right\}_{a}$ for $a\in\overline{1,8}$ with fixed $q_2$ and no constraint on $q_3$. This is done by distributing the color components $Q^a$ on an $8$-dimensional hypersphere of radius given by $q_2$ and then extracting $q_3$ according to Eq.~\eqref{eq:q23}. The resulting \texttt{KDE} built using all numerical values of $q_3$ is shown in the inset plots from Fig.~\ref{fig:mombroadq3dep} and reveals that most classical color charges have a null cubic Casimir $q_3=0$. Using a discrete subset of $q_3$ values, we study the effect of varying $q_3$ by keeping $q_2$ fixed, for both $q_2=4$ and $q_2=4/3$. In both cases, the momentum broadening components show a very weak dependence on $q_3$, with the longitudinal component being more affected at later $\Delta\tau$ times. Lastly, we indirectly test the validity of the Casimir scaling property from Eq.~\eqref{eq:casscal}. In case of Casimir scaling, the ratio $\langle\delta p^2\rangle/D_F$ for $q_2=D_F C_2(F)$ should directly match $\langle\delta p^2\rangle$ obtained with $q_2=C_2$. Nevertheless, our numerical results reveal a weak deviation from Casimir scaling for $\langle\delta p^2_L\rangle$. Interestingly, we separately checked that, as the heavy quark dynamics is further away from the infinite mass static quark scenario, this deviation becomes more pronounced. Violations of Casimir scaling have also been reported for jet quarks propagating in-medium as opposed to in-vacuum \cite{Apolinario:2020nyw}.

\subsubsection{Decorrelation width}

As similarly done for the momentum broadening, we study the $q_3$ dependence of the azimuthal $\sigma_{\Delta\phi}$ and rapidity $\sigma_{\Delta\eta}$ correlation widths, for charm quarks initialized with $p_T(\tau_\mathrm{form}=2\,\mathrm{GeV})$. 
The results from Fig.~\ref{fig:decorrq3dep} show the $\Delta\tau$ evolution of these widths for $q_2=4$ and $q_2=4/3$, by varying $q_3$ with the same set of discrete values used for the momentum broadening. As for $\langle \delta p^2\rangle$, the widths $\sigma_{\Delta\eta, \Delta\phi}$ exhibit a weak dependence on $q_3$, with the exception of $\sigma_{\Delta\eta}$ for $q_2=4$ at larger $\Delta\tau$ values. Nevertheless, such deviations are not problematic since at such late times, the glasma picture becomes unreliable. 

\bibliographystyle{JHEP}
\bibliography{refs}

\end{document}